\documentclass[twocolappendix]{emulateapj}

\defcitealias{kim17}{Paper I}

\usepackage{amsmath}
\usepackage{color}
\usepackage{ulem}
\usepackage{bold-extra}

\usepackage{natbib}

\definecolor{citecolor}{rgb}{0,0,0.5}
\definecolor{linkcolor}{rgb}{1,0,1}
\usepackage[backref,breaklinks,colorlinks,citecolor=blue,linkcolor=magenta]{hyperref}
\usepackage[all]{hypcap}

\slugcomment{Accepted to ApJ}

\newcommand{\software}[1]{{\vskip6pt{\it Software:} #1}}%

\newcommand{\Athena}{{\textit{Athena}}\ignorespaces}

\newcommand\pc{\,{\rm pc}}
\newcommand\kpc{\,{\rm kpc}}
\newcommand\second{\,{\rm s}}
\newcommand\Myr{\,{\rm Myr}}
\newcommand\Gyr{\,{\rm Gyr}}
\newcommand\cm{\,{\rm cm}}
\newcommand\kms{\,{\rm km}\,{\rm s}^{-1}}
\newcommand\Msun{\,M_{\odot}}
\newcommand\Lsun{\,L_{\odot}}
\newcommand\Kel{\,{\rm K}}
\newcommand\kB{{\,k_{\rm B}}}
\newcommand\eV{{\,{\rm eV}}}
\newcommand\Sunit{\,M_{\odot}\,{\rm pc^{-2}}}

\newcommand{\HII}{\ion{H}{2}\ignorespaces}

\newcommand{\xn}{x_{\rm n}}
\newcommand{\alphaB}{\alpha_{\rm B}}
\newcommand{\Mcl}{M_{\rm 0}}
\newcommand{\Rcl}{R_{\rm 0}}
\newcommand{\Sigmacl}{\Sigma_{\rm 0}}

\newcommand{\tffo}{t_{\rm ff,0}}
\newcommand{\vesc}{v_{\rm esc,0}}
\newcommand{\alphaviro}{\alpha_{\rm vir,0}}

\newcommand{\Qi}{Q_{\rm i}}
\newcommand{\ci}{c_{\rm i}}
\newcommand{\Qieff}{Q_{\rm i,eff}}
\newcommand{\Ai}{A_{\rm i}}
\newcommand{\Hi}{H_{\rm i}}
\newcommand{\nion}{n_{\rm i}}
\newcommand{\muH}{\mu_{\rm H}}
\newcommand{\dotNe}{\dot{N}_{\rm e}}
\newcommand{\dotMion}{\dot{M}_{\rm ion}}
\newcommand{\Mion}{M_{\rm ion}}
\newcommand{\tion}{t_{\rm ion}}
\newcommand{\Ne}{N_{\rm e}}
\newcommand{\phiion}{\phi_{\rm ion}}
\newcommand{\phit}{\phi_{t}}
\newcommand{\Sigmaion}{\Sigma_{\rm ion}}

\newcommand{\SFE}{\varepsilon_{*}}
\newcommand{\epstu}{\varepsilon_{\rm ej,turb}}
\newcommand{\epsion}{\varepsilon_{\rm ion}}
\newcommand{\epsej}{\varepsilon_{\rm ej}}
\newcommand{\epsejneu}{\varepsilon_{\rm ej,neu}}
\newcommand{\epsejion}{\varepsilon_{\rm ej,ion}}
\newcommand{\pyield}{p_*/m_*}

\begin{document}


\title{Modeling UV Radiation Feedback from Massive Stars: II.
  Dispersal of Star-Forming Giant Molecular Clouds by Photoionization
  and Radiation Pressure}

\shorttitle{Dispersal of GMCs by UV Radiation Feedback} %
\shortauthors{Kim, Kim, \& Ostriker} %

\author{Jeong-Gyu Kim\altaffilmark{1,2}, Woong-Tae
  Kim\altaffilmark{1,2}, \& Eve C.~Ostriker\altaffilmark{2}}

\affil{$^1$Department of Physics \& Astronomy, Seoul National University, Seoul 08826, Republic of Korea}
\affil{$^2$Department of Astrophysical Sciences, Princeton University, Princeton, NJ 08544, USA}
\email{jgkim@astro.snu.ac.kr, wkim@astro.snu.ac.kr, eco@astro.princeton.edu, skinner15@llnl.gov}






\begin{abstract}
  UV radiation feedback from young massive stars plays a key role in
  the evolution of giant molecular clouds (GMCs) by photoevaporating
  and ejecting the surrounding gas. We conduct a suite of radiation
  hydrodynamic simulations of star cluster formation in
  marginally-bound, turbulent GMCs, focusing on the effects of
  photoionization and radiation pressure on regulating the net star
  formation efficiency (SFE) and cloud lifetime. We find that the net
  SFE depends primarily on the initial gas surface density,
  $\Sigma_0$, such that the SFE increases from 4\% to 51\% as
  $\Sigma_0$ increases from $13\,M_{\odot}\,{\rm pc}^{-2}$ to
  $1300\,M_{\odot}\,{\rm pc}^{-2}$. Cloud destruction occurs within
  $2$--$10\,{\rm Myr}$ after the onset of radiation feedback, or
  within $0.6$--$4.1$ freefall times (increasing with $\Sigma_0$).
  Photoevaporation dominates the mass loss in massive, low
  surface-density clouds, but because most photons are absorbed in an
  ionization-bounded Str\"{o}mgren volume the photoevaporated gas
  fraction is proportional to the square root of the SFE. The measured
  momentum injection due to thermal and radiation pressure forces is
  proportional to $\Sigma_0^{-0.74}$, and the ejection of neutrals
  substantially contributes to the disruption of low-mass and/or
  high-surface density clouds. We present semi-analytic models for
  cloud dispersal mediated by photoevaporation and by dynamical mass
  ejection, and show that the predicted net SFE and mass loss
  efficiencies are consistent with the results of our numerical
  simulations.
\end{abstract}


\section{INTRODUCTION}

Ultraviolet (UV) radiation produced by newborn star clusters
profoundly affects the evolution of giant molecular clouds (GMCs),
where most star formation in the local Universe takes place. The UV
photons dissociate and ionize cold molecular gas that could otherwise
fuel further star formation \citep[e.g.,][]{whi79}. Photoionization
increases the thermal pressure within GMCs by three orders of
magnitude, and expansion of high-pressure ionized bubbles (\HII\
regions) can not only mechanically unbind the parent cloud but also
induce turbulent motions in the surrounding interstellar medium (ISM)
\citep{gri09,wal12,med14}. In addition to photoionization, UV photons
exert radiation pressure on dust, which is collisionally coupled to
the gas. This radiation pressure force alters the internal structure
of \HII\ regions surrounding newborn luminous clusters \citep{dra11}
and helps to expel the gas from GMCs. While the above processes help
to quench star formation locally and globally, it has also been
proposed that UV radiation can stimulate star formation either in the
``collect and collapse'' scenario in which swept-up shells around
\HII\ regions grow in mass and gravitationally collapse
\citep[e.g.,][]{elm77,whi94,hos06,dal07,deh10} or via
``radiation-driven implosion'' in which an ionization front drives
converging shock waves toward the centers of pre-existing globules
\citep[e.g.,][]{ber89,sug89,lef94,bis11}. Still, although localized
``triggering'' may occur, existing simulations (see below) show that
UV radiation feedback overall has a negative impact on star formation.

Compared to what would be expected from unimpeded freefall collapse,
star formation in GMCs is empirically a rather slow and inefficient
process \citep[e.g.][]{zuc74a,zuc74b,eva91}. While the typical GMC
freefall time is only of order $\sim 1$--$10 \Myr$, the gas depletion
timescale $t_{\rm dep}$ estimated from the ratio between gas mass and
star formation rate is $\sim 1\Gyr$ for molecular gas averaged over
$\sim \kpc$ scales in normal disk galaxies
\citep[e.g.,][]{big08,ler13} and somewhat shorter in galactic centers
and starburst environments \citep[e.g.,][]{dad10,ler15,per16} as well
as individual molecular clouds \citep[e.g.,][]{ken12,vut16}. In
Galactic star-forming clouds, the observed star formation efficiency
(SFE)---defined as the ratio of stellar to gas mass--- varies widely,
ranging over $\sim 0.0001$--$0.3$, and is only a few percent on
average
\citep[e.g.,][]{mye86,moo88,wil97,car00,eva09,mur10,gar14,lee16,vut16},
while evidence is accumulating that clouds in extremely dense
environments may have higher SFEs
\citep[e.g.,][]{mei02,tur15,con16,tur17}. Yet, it is still uncertain
what physical processes are responsible for the diversities in
observed SFEs.

A related issue concerns the lifetime of clouds. For an idealized
isolated cloud with steady star formation, the cloud lifetime
$t_{\rm cl}$ would be related to the depletion timescale via
$t_{\rm cl} = \SFE t_{\rm dep}$, where the net SFE, $\SFE$, is the
fraction of the initial cloud mass that will \emph{ever} become stars.
No consensus exists over how long an individual molecular cloud
survives as a coherent entity \citep{hey15}. On the one hand, it has
been argued that the short age spreads ($\lesssim 5\Myr$) of stellar
populations in nearby star-forming clouds supports the picture that
star-forming clouds are dispersed rapidly---on the cloud's dynamical
timescale (e.g. \citealt{har01a,har01b}, but see also \citealt{dar14},
who argue that cluster formation takes several local dynamical
timescales). On the other hand, estimates based on the fraction of
GMCs that are spatially coincident with \HII\ regions and young star
clusters in nearby galaxies favor lifetimes of a few tens of Myrs
(\citealt{eng03,bli07,kaw09,miu12}; see also
\citealt{mur11,mei15,lee16}). The cloud lifetime also has implications
for the origin and maintenance of turbulence in GMCs. While supersonic
turbulence appears to be pervasive in GMCs \citep{elm04}, its energy
begins to decay within one large-scale crossing time in the absence of
driving \citep[e.g.,][]{mac98,sto98}. If GMCs are very long lived,
turbulence must be driven continuously (but not excessively) by
internal or external processes to support them against gravitational
collapse without dispersal (\citealt{kru06,gol11}), while this kind of
finely calibrated turbulent driving is not needed if GMCs are
dispersed rapidly \citep{elm00,bal06,bal11}.

A number of theoretical studies have invoked the destructive role of
UV radiation feedback to explain the observed SFEs and lifetimes of
GMCs \citep[see reviews by][]{mck07,kru14}. Analytic models based on
idealized solutions for expanding blister-type \HII\ regions found
that the net SFE of $\SFE \sim 5$--$15\%$ is sufficient to disperse
typical GMCs in the Milky Way (mass $M \sim$ $10^5$--$10^6 \Msun$ and
surface density $\Sigma \sim 10^2 \Sunit$) by photoevaporation and
dynamical disruption in a few tens of ${\rm Myr}$
\citep{whi79,wil97,mat02,kru06}. \citet{kru09}, \citet{fal10}, and
\citet{mur10} highlighted the importance of radiation pressure on dust
grains in controlling the dynamics of \HII\ regions in dense, massive
star-forming environments. \citet{kim16} studied dynamical disruption
of clouds by considering the expansion of a swept-up spherical shell
surrounding a central \HII\ region, and found that the minimum SFE
required for cloud disruption increases primarily with the cloud's
surface density and that the disruption timescale is comparable to the
freefall time. These models also suggest that radiation pressure is
expected to be more important than ionized-gas pressure in massive and
high surface density clouds \citep[see also][]{kru09,fal10}. More
recently, \citet{rah17} developed a semi-analytic model for the
dynamics of an expanding shell formed around a star cluster including
the effects of radiation pressure, stellar winds, and supernovae (SNe)
(but not including effects of ionized-gas pressure), and evaluated the
minimum SFE as well as the relative importance of each feedback
mechanism.

While analytic approaches offer valuable insights into the physical
processes involved in cloud dispersal, they are limited to smooth
and/or spherically symmetric density distributions. Real GMCs are
turbulent and extremely inhomogeneous. As a consequence, shell
expansion induced by \HII\ regions from multiple subclusters is not
spherically symmetric, and neither embedded nor blister \HII\ regions
present smooth interior surfaces to photoionizing radiation. To
quantitatively follow the formation and evolution of multiple
cluster-containing \HII\ regions in highly turbulent, inhomogeneous
clouds, and to quantitatively assess the consequences of the complex
interplay between gas and UV radiation, numerical radiation
hydrodynamic (RHD) simulations are necessary \citep[see recent reviews
by][]{kru14,dal15}.

Several numerical studies have investigated the effects of
photoionization feedback on dynamical evolution of GMCs. For instance,
\citet{vaz10} and \citet{col13} allowed for photoheating by ionizing
radiation in the evolution of GMCs produced by colliding flows and
showed that the feedback reduces SFEs significantly. Using smoothed
particle hydrodynamics simulations, \citet{dal12,dal13} performed RHD
simulations of star cluster formation in turbulent GMCs over a $3\Myr$
timescale before the first supernova event occurs. Their parameter
study showed that while photoionization can unbind a significant
fraction of material in low-mass, diffuse clouds, it has a limited
impact for clouds with escape velocities larger than the sound speed
of ionized gas. \citet{gav17} studied the early (2 Myr) dynamical
evolution of a star cluster formed in a turbulent cloud with mass
$2.5 \times 10^4\Msun$ and surface density $250 \Sunit$ and found that
photoionization can limit the net SFE to $20\%$. \citet{gee17}
performed RHD simulations of magnetized, $10^4 \Msun$-mass clouds of
varying size and found that the net SFE increases from $0.04$ to $0.6$
as the surface density increases from $14$ to $1100 \Sunit$.

Considering the effects of non-ionizing UV, \citet{ras16} conducted
RHD simulations of cluster-forming turbulent clouds, focusing
exclusively on the radiation pressure from singly-scattered,
non-ionizing UV. They found that the resulting net SFE
($\SFE \sim 0.1$--$0.6$ for $\Sigma \sim 10$--$300 \Sunit$) in
turbulent cloud simulations is much higher than that expected for the
case of a uniform shell. This is because turbulent shock compression
leads to a broad (log-normal) distribution of the gas surface density,
which in turn increases the fraction of super-Eddington, high
surface-density gas that is difficult to unbind and hence subject to
star formation \citep[see also][]{tho16}. \citet{ras16} concluded that
feedback from radiation pressure alone is unable to explain the low
SFEs of observed GMCs.

More recently, \citet{how17} performed a series of RHD simulations to
study the early phase ($\lesssim 5\Myr$) of GMC evolution, including
both photoionization and radiation pressure. They considered a set of
GMC models with the same mean density ($n_{\rm H} \sim 150 \cm^{-3}$)
but differing mass ($10^4$--$10^6 \Msun$), and found a modest degree
of suppression in the SFE compared to the runs in which they turned
off radiation feedback. In their models, the impact of radiative
feedback varied without a clear trend with the cloud mass; only
intermediate-mass clouds ($5\times 10^4\Msun$ and $10^5 \Msun$) were
fully ionized and destroyed in $5\Myr$.

While the above numerical studies have improved our understanding of
the effects of UV radiation on GMC evolution and star formation
regulation, several important issues still need to be addressed.
First, although several studies have emphasized the importance of
photoionization and radiation pressure for cloud dispersal, as yet
there is no systematic accounting that quantifies to what extent each
mechanism, alone and combined, is responsible for the net SFE and
cloud disruption. Second, most of the previous simulations have only
used approximate methods for treating the radiative transfer problem
for multiple point sources. For example, \citet{vaz10} incorporated
photoionization feedback by depositing thermal energy to a gas cell
where a stellar particle resides. \citet{dal12,dal13} adopted the
``Str\"{o}mgren volume technique'' to calculate the ionization state
of gas, but did not consider the effects of radiation pressure on
dust. Simulations by \citet{ras16,ras17}, \citet{gav17}, and
\citet{gee17} adopted two-moment methods to evolve the radiation
energy and momentum on a grid with the $M_1$-closure. Although the
$M_1$-closure is accurate for a spatially concentrated source
distribution, it becomes less accurate when multiple point sources are
widely distributed; this method also has inherently limited resolution
in the immediate vicinity of point sources \citep[see][hereafter Paper
I]{kim17}. Thus, a firm quantitative assessment of the importance of
both photoionization and radiation pressure on cloud dispersal
requires a more accurate solution of the radiative transfer equation
than has previously been available.

In this work, we have carried out RHD simulations of star cluster
formation in turbulent clouds employing the adaptive ray tracing
method \citep{abe02}, which enables us to accurately solve the
radiative transfer problem for multiple point sources and both
ionizing and non-ionizing radiation. In \citetalias{kim17} we
described our implementation and tests of adaptive ray tracing in the
\Athena\ magnetohydrodynamics code \citep{sto09}, for which we adopted
the novel parallelization algorithm proposed by \citet{ros17} as well
as several other improvements. The excellent parallel performance this
has enabled for ray tracing allows us to run a large number of
simulations, probing a range of cloud masses and sizes efficiently. In
the present paper, our primary goals are (1) to quantify the
dependence on the cloud properties of the net SFE, timescale for cloud
disruption, mass of ionized outflows, and momentum transferred to gas
outflows, and (2) to assess the relative importance of photoionization
and radiation pressure in various environments. In addition, we
develop analytic predictions for mass loss based on the physical
scalings for photoevaporation and momentum injection, and we compare
the predictions for the net SFE with the numerical results.

The organization of this paper is as follows. Section~\ref{s:method}
presents the numerical methods and cloud parameters that we adopt.
Section~\ref{s:overview} first describes the overall evolution of our
fiducial model, and then explores the parameter dependence of various
integrated physical quantities for different models. In
Section~\ref{s:ML}, we present a detailed analysis of the mass loss
caused by photoevaporation and momentum injection. In
Section~\ref{s:dispersal}, we construct the semi-analytic models for
the net SFE and mass loss efficiencies of GMCs regulated by UV
radiation feedback, and compare the model predictions with the
numerical results. In Section~\ref{s:sum}, we summarize and discuss
the astronomical implications of our results. In
Appendix~\ref{s:app1}, we present the results of convergence study for
the fiducial model. In Appendix~\ref{s:app2}, we compare our numerical
results for SFE in radiation-pressure only models to analytic
predictions.

\section{NUMERICAL METHODS}\label{s:method}

For our numerical simulations, we use the Eulerian grid-based code
\Athena\ \citep{sto08} equipped with additional physics modules for
self-gravity, sink particles, and radiative transfer from point
sources. In \citetalias{kim17}, we presented a detailed description of
our implementation of the adaptive ray tracing algorithm for multiple
point sources in \Athena, including parallelization and the methods
for solving photoionization and recombination. \citetalias{kim17} also
described the initial conditions and problem setup for our simulations
of cluster-forming turbulent clouds. Below we briefly summarize the
highlights of our numerical methods and describe the initial cloud
parameters.

\subsection{Radiation Hydrodynamics Scheme}\label{s:RHScheme}

For hydrodynamics, we employ \Athena's van Leer-type time
integrator \citep{sto09}, the HLLC Riemann solver, and a piecewise
linear spatial reconstruction scheme. We use the fast Fourier
transformation Poisson solver with open boundary conditions
\citep{ski15} to calculate the gravitational forces from gas and
stars; the stellar contribution is handled with the particle-mesh
method using the triangular-shaped-cloud interpolation scheme
\citep{hoc81}.

We apply the sink particle technique of \citet{gon13} to model the
formation and growth of star clusters. We create a star particle if a
cell with density above the Larson-Penston threshold density
($\rho_{\rm crit}\equiv 8.86c_{\rm s, neu}^2/(\pi G \Delta x^2)$ for
the sound speed of neutral gas $c_{\rm s,neu}$ and the grid spacing
$\Delta x$) has a converging velocity field and corresponds to the
local minimum of the gravitational potential. The accretion rates of
mass and momentum onto each sink are computed from the flux through
the surfaces of the $3^3$-cell control volume centered at the sink
particle; this control volume acts like internal ghost zones within
the simulation domain. The positions and velocities of sink particles
are updated using a leapfrog integrator. We allow sink particles to
merge if their control volumes overlap.

Due to limited resolution, the sink particles in our simulations
represent subclusters rather than individual stars\footnote{The
  typical mass of newly formed sink particles is
  $M_{\rm sink} \sim 10 \rho_{\rm crit}\Delta x^3 \approx 240 \Msun
  (\Rcl/20\pc)(N_x/256)^{-1}$, where $\rho_{\rm crit}$ is the
  threshold for sink particle creation, $\Rcl$ is the initial cloud
  radius, and $N_x$ is the number of grid zones in the domain in one
  direction.}, and may not fully sample the initial mass function.
Using the mass-dependent light-to-mass ratios from \citet{kim16}, we
determine the total UV luminosity of cluster particles for two
frequency bins, $L_{\rm i}$ and $L_{\rm n}$, representing Lyman
continuum photons and far-UV photons, such as
$L \equiv L_{\rm i} + L_{\rm n} = \Psi M_*$ and
$\Qi \equiv L_{\rm i}/(h\nu_{\rm i}) = \Xi M_*$, where $M_*$ is the
total cluster mass,
$\Psi = 10^{2.98 \mathcal{X}^6/(29.0 + \mathcal{X}^6)} \Lsun
\Msun^{-1}$,
$\Xi = 10^{46.7\mathcal{X}^7/(7.28 + \mathcal{X}^7)}
\second^{-1}\Msun^{-1}$ with
$\mathcal{X} = \log_{10} (M_*/M_{\odot})$, and $h\nu_{\rm i} = 18 \eV$
is the mean energy of ionizing photons. Note that
$\Psi\simeq 943 \Lsun \Msun^{-1}$ and
$\Xi \simeq 5.05\times 10^{46}\second^{-1}\Msun^{-1} $ are almost
constant for $M_*\gtrsim 10^3\Msun$, while varying steeply for
$M_* <10^3\Msun$ (see Figure 14 of \citealt{kim16}). The luminosity of
individual cluster particles are assumed to be proportional to their
mass. The use of mass-dependent conversion factors approximately
captures the effect of undersampling at the massive end of the initial
mass function when the total cluster mass is below $\sim 10^3 \Msun$.

In the present paper, we do not consider evolution of the stellar
light-to-mass ratio. The UV luminosity of a coeval stellar population
that well samples the initial mass function remains approximately
constant for the main sequence lifetime of the most massive star,
implying that ionizing and non-ionizing luminosities would in practice
decay rapidly after timescales of $3\Myr$ and $8\Myr$, respectively
\citep[e.g.,][]{par03}. Caveats related to our adoption of
age-independent luminosity are discussed in Section~\ref{s:discuss}.

The radiative transfer of ionizing and non-ionizing photons is handled
by the adaptive ray tracing method \citep{abe02}. For each radiating
star particle, we generate $12 \times 4^4 = 3072$ initial rays and
allow photon packets to propagate radially outward, computing the
cell-ray intersection length and the corresponding optical depth on a
cell-by-cell basis. The volume-averaged radiation energy and flux
densities of a cell are computed from the sum of the contributions
from all rays that pass through the cell (see Section 2.1 of
\citetalias{kim17}). These averages are used to compute the radiation
force from the combined non-ionizing and ionizing radiation fields,
and the ionization rate of neutral gas. Because the fluid variables
and gravity in the $3^3$ cells surrounding each sink/source particle
are unresolved, we do not allow photon packets to interact with the
gas within these control volumes. The radiation field directions are
angularly well resolved, and the rays contain all of the photon
packets from each source, when they emerge from the control volume
around each source particle. This eliminates the potential problem of
momentum cancellation caused by volume-averaging in the cell
containing a source, noted recently by \citet{hop18}, without the need
to introduce assumptions regarding the sub-cell distributions of gas
density, gravity, or the radiation field within individual zones that
contain sink/source particles.

The rays are discretized and split based on the HEALPix framework
\citep{gor05}, ensuring that each grid cell is sampled by at least
four rays per source. We rotate ray propagation directions randomly at
every hydrodynamic time step to reduce the numerical errors due to
angle discretizations \citep{kru07}. For simplicity, we use constant
photoionization cross-section for neutral hydrogen atom
$\sigma_{\rm H^0} = 6.3 \times 10^{-18} \cm^{-2}$
\citep{kru07}\footnote{Our adopted photoionization cross-section
  corresponds to the value at the Lyman edge. The frequency-averaged
  cross-section depends on the local radiation spectrum and is
  typically smaller by a factor of $\sim 2$ \citep[e.g.,][]{bac15}.
  Although the use of smaller $\sigma_{\rm H^0}$ can increase the
  neutral fraction of ionized gas inside the \HII\ region, we have
  checked that it has little impact on the simulation outcome.} and
constant dust absorption (and pressure) cross-section per H nucleon
$\sigma_{\rm d} = 1.17 \times 10^{-21} \cm^{-2}\,{\rm H}^{-1}$ both
for ionizing and non-ionizing radiation \citep[e.g.,][]{dra11}. We
determine the degree of hydrogen ionization by solving the
time-dependent rate equation for hydrogen photoionization and
recombination (under the on-the-spot approximation). The ray-trace and
ionization update are subcycled relative to the hydrodynamic update.
The timestep size for each substep is determined to ensure that the
maximum change in the neutral fraction $\xn = n_{\rm H^0}/n_{\rm H}$
is less than 0.1. The source term updates for the ionization fraction
and radiative force are explicit and performed at every substep in an
operator-split fashion. The gas temperature is set to vary according
to $\xn$ between $T_{\rm neu} = 20 \Kel$ and $T_{\rm ion}=8000 \Kel$,
the temperature of the fully neutral and ionized gas, respectively.
The corresponding isothermal sound speeds are
$c_{\rm s,neu} = 0.26 \kms$ and $c_{\rm s,ion} = 10.0 \kms$.

We note that determining gas temperature solely based on the neutral
fraction, together with adoption of constant cross-sections for
photoionization and dust absorption, will certainly simplify
temperature distributions inside \HII\ regions compared to what would
be more complex in realistic environments. To properly model
ionization and temperature structure, as well as small-scale dynamical
instabilities of ionization fronts that may operate
\citep[e.g.,][]{kim14,bac15b}, one has to accurately solve the energy
equation after considering various cooling and heating processes as
well as the frequency dependence of the cross-sections.

\subsection{Initial and Boundary Conditions}

Our initialization and boundary treatment are the same as in
\citet{ski15} and \citet{ras16}. The initial conditions for each model
consist of an isolated, uniform-density sphere of neutral gas with
mass $\Mcl$ and radius $\Rcl$ surrounded by a tenuous external medium
situated in a cubic box with sides
$L_{\rm box}=4\Rcl$.\footnote{Including the tenuous external medium,
  the initial gas mass in our simulation domain is 1.015$\Mcl$.} The
cloud is initially supplied with turbulent energy with velocity power
$|\delta \mathbf{v}_k|^2 \propto k^{-4}$ for
$2 \le kL_{\rm box}/(2\pi) \le 64 $ and zero power otherwise. We
adjust the amplitude of the velocity perturbations to make the cloud
marginally bound gravitationally, with the initial virial parameter of
$\alphaviro = 5\sigma_0^2 \Rcl/(3G\Mcl) = 2$, where $\sigma_0$ denotes
the initial (three-dimensional) velocity dispersion. There is no
subsequent artificial turbulent driving. The standard resolution for
the simulation domain is $N_{\rm cell}=256^3$ cells.

We apply diode-like outflow boundary conditions both at the boundaries
of the simulation box and at the boundary faces of the control volume
of each sink particle. The diode-like conditions allow mass and
momentum to flow into ghost zones but not out from it. When sink
particles accrete while moving across grid zones, we ensure that the
total mass and momentum of gas and stars are conserved by taking into
account differences in the mass and momentum of cells entering and
leaving the control volume. Over the course of evolution, we
separately monitor the mass outflow rates of neutral and ionized gas
from the computational domain, as well as escape fractions of ionizing
and non-ionizing photons.

\begin{figure}
  \epsscale{1.25}\plotone{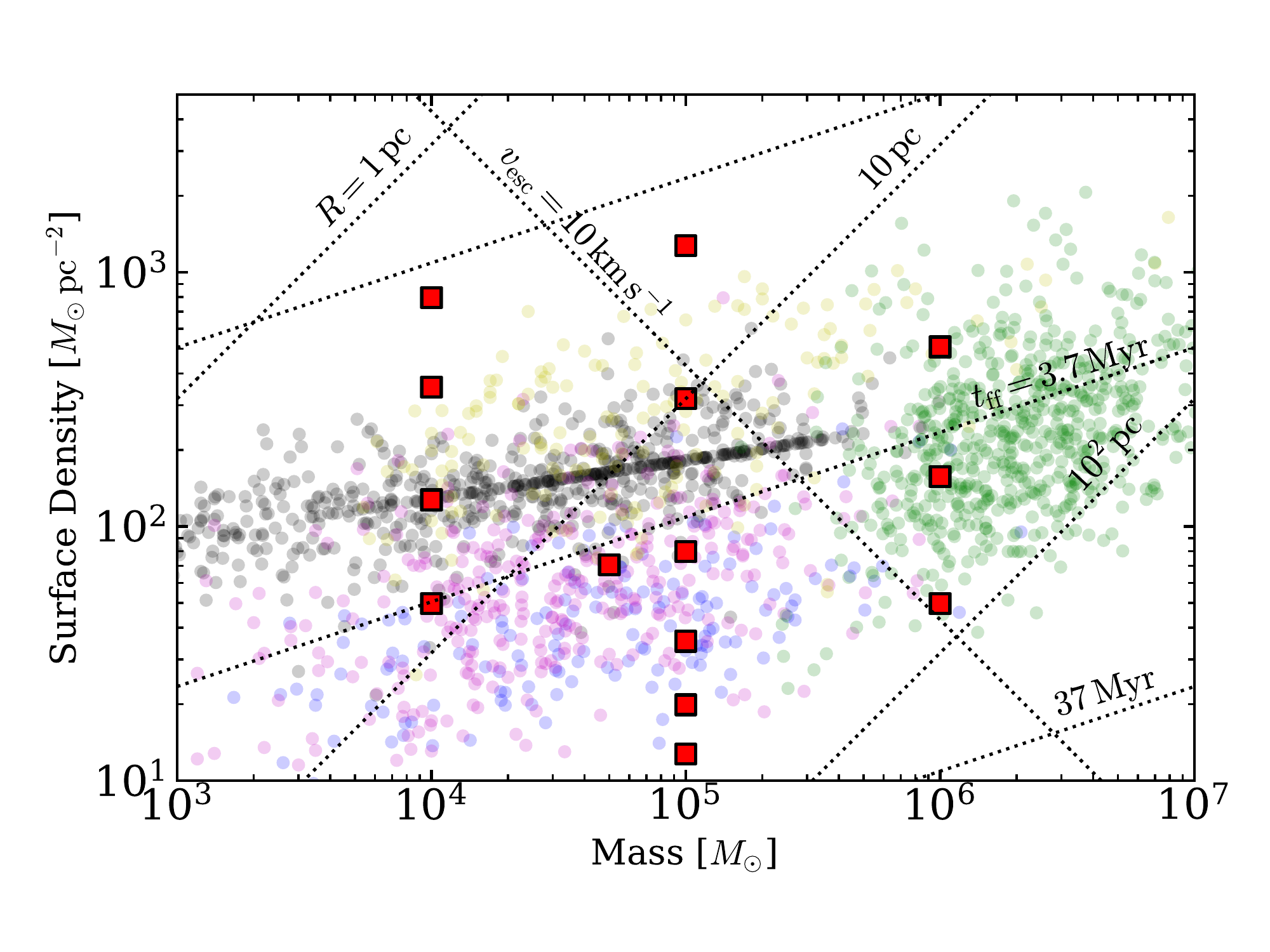}
  \caption{Locations of our model clouds (red squares) in the
    $M$--$\Sigma$ plane. Other symbols denote the observed molecular
    clouds in the literature: ${}^{13}{\rm CO}$ Galactic Ring Survey
    from \citet{hey09} (blue) and \citet{rom10} (black),
    ${}^{12}{\rm CO}$ survey of Galactic Center \citet{oka01}
    (yellow), ${}^{12}{\rm CO}$ survey of Large Magellanic Clouds from
    \citet{won11} (magenta), and ${}^{12}{\rm CO}$ survey of M51 from
    \citet{col14} (green). Dotted lines draw the loci of constant
    radius, escape velocity, and freefall time.}\label{f:cloud}
\end{figure}

\subsection{Models}

Our aim is to explore the effects of radiation feedback on cloud
dispersal across a range of cloud masses and sizes, corresponding to
GMCs observed in the Milky Way and nearby galaxies. These clouds are
usually optically thick to UV radiation
($\Sigma \gtrsim \kappa_{\rm UV}^{-1} \sim 10^1 \Sunit$) but
transparent to dust-reprocessed infrared radiation
($\Sigma \lesssim \kappa_{\rm IR}^{-1} \sim 10^3 \Sunit $). The
momentum deposition by multiple scatterings of infrared photons may
play an important role in the dynamics of extremely dense and massive
clouds, such as the progenitors of super star clusters, in dust-rich
environments \citep[e.g.,][]{ski15,tsa17}.

\capstartfalse
\begin{deluxetable}{lcccccc}
\tabletypesize{\footnotesize}
\tablewidth{0pt}\tablecaption{Model Cloud Parameters}
\tablehead{
\colhead{Model} &
\colhead{$\Mcl$} &
\colhead{$\Rcl$} &
\colhead{$\Sigmacl$} &
\colhead{$n_{{\rm H,0}}$} &
\colhead{$t_{{\rm ff,0}}$} &
\colhead{$v_{\rm esc,0}$} \\
\colhead{(1)} &
\colhead{(2)} &
\colhead{(3)} &
\colhead{(4)} &
\colhead{(5)} &
\colhead{(6)} &
\colhead{(7)}}
\startdata
{\tt M1E5R50} & $1 \times 10^5$ & $50.0$ & $12.7$ & $ 5.5$ & $18.5$ & $ 4.1$ \\
{\tt M1E5R40} & $1 \times 10^5$ & $40.0$ & $19.9$ & $10.8$ & $13.2$ & $ 4.6$ \\
{\tt M1E5R30} & $1 \times 10^5$ & $30.0$ & $35.4$ & $25.5$ & $ 8.6$ & $ 5.4$ \\
{\tt M1E4R08} & $1 \times 10^4$ & $ 8.0$ & $49.7$ & $134.7$ & $ 3.7$ & $ 3.3$ \\
{\tt M1E6R80} & $1 \times 10^6$ & $80.0$ & $49.7$ & $13.5$ & $11.8$ & $10.4$ \\
{\tt M5E4R15} & $5 \times 10^4$ & $15.0$ & $70.7$ & $102.2$ & $ 4.3$ & $ 5.4$ \\
{\tt \textbf{M1E5R20}} & $\bf 1 \times 10^5$ & $\bf 20.0$ & $\bf 79.6$ & $\bf 86.2$ & $\bf 4.7$ & $\bf 6.6$ \\
{\tt M1E4R05} & $1 \times 10^4$ & $ 5.0$ & $127.3$ & $551.8$ & $ 1.9$ & $ 4.1$ \\
{\tt M1E6R45} & $1 \times 10^6$ & $45.0$ & $157.2$ & $75.7$ & $ 5.0$ & $13.8$ \\
{\tt M1E5R10} & $1 \times 10^5$ & $10.0$ & $318.3$ & $689.7$ & $ 1.7$ & $ 9.3$ \\
{\tt M1E4R03} & $1 \times 10^4$ & $ 3.0$ & $353.7$ & $2554.6$ & $ 0.9$ & $ 5.4$ \\
{\tt M1E6R25} & $1 \times 10^6$ & $25.0$ & $509.3$ & $441.4$ & $ 2.1$ & $18.6$ \\
{\tt M1E4R02} & $1 \times 10^4$ & $ 2.0$ & $795.8$ & $8621.6$ & $ 0.5$ & $ 6.6$ \\
{\tt M1E5R05} & $1 \times 10^5$ & $ 5.0$ & $1273.2$ & $5517.8$ & $ 0.6$ & $13.1$ \\
\enddata
\tablecomments{Parameters listed are for initial conditions of
  spherical clouds. Column 1: model name. Column 2: cloud mass
  ($M_{\odot}$). Column 3: cloud radius (pc). Column 4: gas surface
  density $(M_{\odot}\,{\rm pc}^{-2})$. Column 5: number density of
  neutral gas ($\rm cm^{-3}$). Column 6: freefall time (Myr). Column
  7: escape velocity at the cloud surface ($\rm km\,s^{-1}$). The
  fiducial model {\tt M1E5R20} is shown in bold.}\label{t:model}
\end{deluxetable}
\capstarttrue

We consider 14 clouds with $\Mcl$ in the range of $10^4 \Msun$ to
$10^6 \Msun$ and $\Rcl$ from $2\pc$ to $80\pc$: the resulting initial
surface density $\Sigmacl = \Mcl/(\pi \Rcl^2)$ is in the range from
$12.7 \Sunit$ to $1.27 \times 10^3 \Sunit$. \autoref{f:cloud} plots
the locations (red squares) of our model clouds in the $M$--$\Sigma$
domain, compared to observed GMCs (circles) compiled from the
literature (note that a range of different methods have been adopted
in the literature to estimate observed cloud surface density and
mass). \autoref{t:model} lists the initial physical parameters of
our model clouds. Column 1 gives the name of each model. Columns 2--4
list $\Mcl$, $\Rcl$, and $\Sigmacl$, respectively. Columns 5 and 6
give the number density of hydrogen atoms $n_{\rm H,0}$ and the
freefall time $\tffo = \sqrt{3\pi/(32 G\rho_0)}$. Finally, Column 7
gives the escape velocity $\vesc=\sqrt{2G\Mcl/\Rcl}$ at the cloud
surface. The initial turbulent velocity dispersion in each cloud is
$\sigma_0 = 0.77 \vesc$. We take model {\tt M1E5R20} with
$\Mcl = 10^5 \Msun$, $\Rcl = 20\pc$, and $\tffo=4.7 \Myr$ as our
fiducial model, which is typical of Galactic GMCs and comparable in
mass and size to the Orion A molecular cloud \citep[e.g.,][]{wil05}.

In addition to the standard runs including both photoionization and
radiation pressure (PH+RP), we run two control simulations for each
cloud, in which either only photoionization (PH-only) or only
radiation pressure (RP-only) is turned on. This enables us to isolate
the effect of each feedback mechanism and thus to indirectly assess
their relative importance in cloud dispersal. For the fiducial cloud,
we additionally run low- and high-resolution models ({\tt
  M1E5R20\_N128} and {\tt M1E5R20\_N512}) with $N_{\rm cell} = 128^3$
and $512^3$ as well as a no-feedback run ({\tt M1E5R20\_nofb}), in the
last of which stellar feedback is turned off. In
Appendix~\ref{s:app1}, we compare the results from the models with
different resolution. Although star formation completes slightly later
in simulations with higher resolution, the overall evolutionary
behavior is qualitatively the same and key quantitative outcomes (such
as the SFE) are quite close at different resolution, following a
converging trend. All simulations are run over $4$--$7$ freefall
times, long enough for star formation to complete and for radiation
feedback to evacuate the remaining gas from the simulation domain.

\section{SIMULATION RESULTS}\label{s:overview}

We begin by presenting the overall temporal evolution of the fiducial
model {\tt M1E5R20} in Section~\ref{s:overall}. Other models exhibit a
similar evolutionary behavior, although there are significant changes
in the net SFE, timescales for star formation and cloud dispersal, and
velocity of outflowing gas. These will be presented in
Section~\ref{s:eps}--\ref{s:vej}.

\subsection{Overall Evolution: Fiducial Model}\label{s:overall}

Similarly to the simulations of \citet{ras16}, the early evolution of
our fiducial model is governed by turbulence and self-gravity. The
density structure becomes increasingly filamentary as a result of
shock compression. The densest regions---which may be at junctions of
filaments, or simply overdense clumps within filaments---undergo
gravitational collapse, leading to the formation of sink particles. In
the fiducial model the first collapse occurs at $t_{*,0}/\tffo=0.4$.

\begin{figure}[!t]
  \epsscale{1.2}\plotone{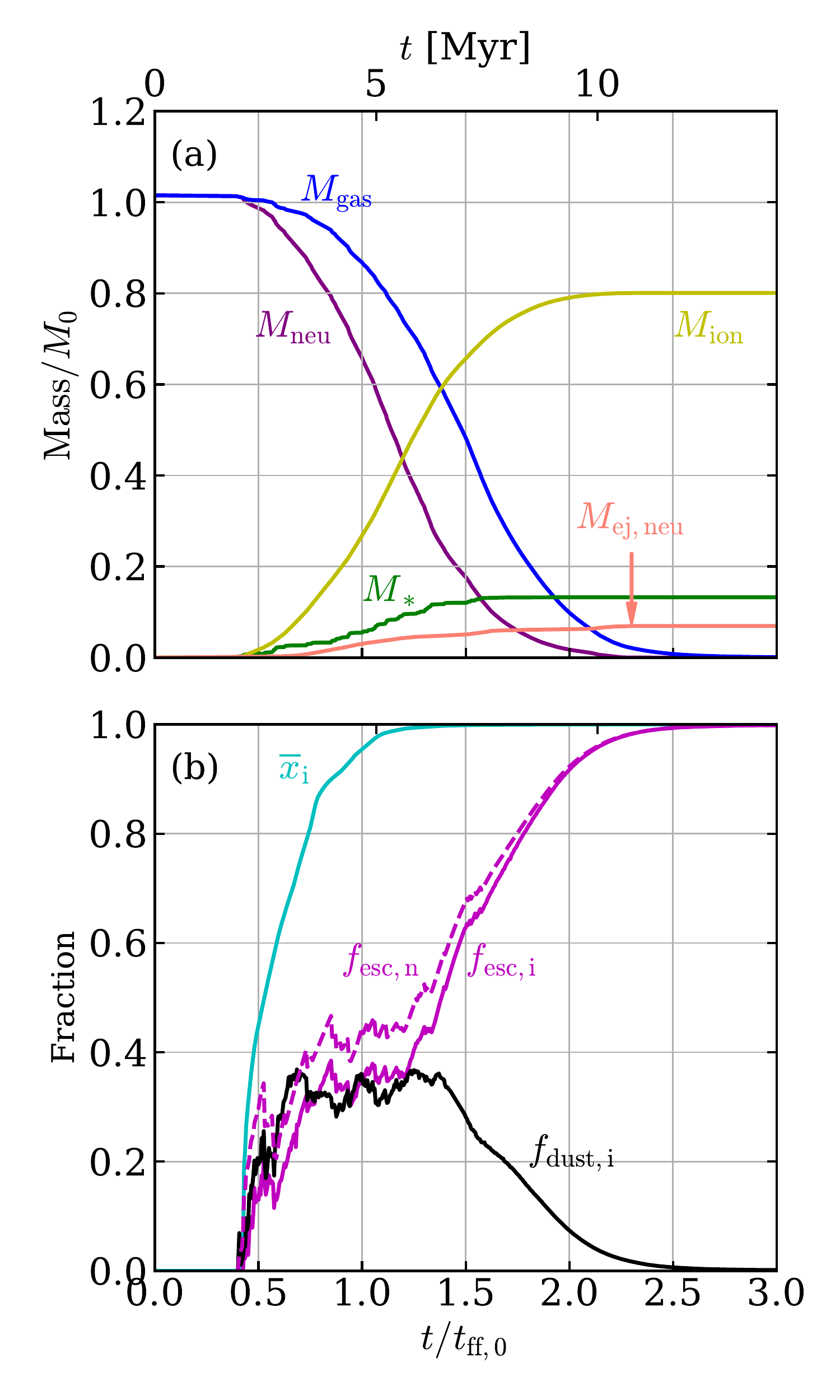}
  \caption{Evolutionary histories of the fiducial model with
    $\Mcl=10^5\Msun$ and $\Rcl = 20\pc$. (a) The total gas mass
    $M_{\rm gas}$ in the simulation volume (blue), the neutral gas
    mass $M_{\rm neu}$ in the simulation volume (purple), the stellar
    mass $M_*$ (green), the ejected neutral gas mass $M_{\rm ej,neu}$
    (salmon), and the mass of the photoevaporated gas $\Mion$
    (yellow). (b) The volume fraction of the ionized gas
    $\overline{x}_{\rm i}$ (cyan), the fraction of ionizing radiation
    absorbed by dust $f_{\rm dust,i}$ (black), and the escape
    fractions of ionizing ($f_{\rm esc,i}$; solid magenta) and
    non-ionizing ($f_{\rm esc,n}$; dashed magenta) radiation are
    plotted as functions of time.}\label{f:hst}
\end{figure}

Due to star formation and the ensuing feedback, the gas in our
simulations is ultimately channelled into three distinct states. One
portion of the gas gravitationally collapses and is accumulated in
sink particles as a total stellar mass $M_*(t)$. Strong ionizing
radiation from newly-formed stars ionizes a portion of the
initially-neutral gas, with the total mass of photoevaporated gas
increasing in time as $M_{\rm ion}(t)$. This includes the ionized gas
currently in the simulation box and the cumulative ionized gas that
has left the simulation domain. Over time, all of the neutral gas in
the domain either collapses to make stars, is photoevaporated and
ejected, or is ejected while still neutral. We denote the cumulative
ejected mass of the neutral gas by $M_{\rm ej,neu}(t)$.\footnote{We
  regard the neutral gas as being ejected if it reaches the outer
  boundary of the simulation box, although one can use a more rigorous
  criterion by testing the gravitational boundedness of individual gas
  parcels \citep[e.g.,][]{dal12}.} Of course, gas that is photoionized
may recombine before it flows out of the box, but at the end of the
simulation when no gas remains in the domain,
$M_{\rm ion,final}=M_{\rm ion}(t_{\rm final})$. The condition of mass
conservation requires
$\Mcl=M_{*,{\rm final}} + M_{\rm ion,final} + M_{\rm ej,neu,final}$ at
the end of the run.

\autoref{f:hst}(a) plots the temporal changes of the total gas mass
$M_{\rm gas}$ in the simulation volume as well as $M_*$,
$M_{\rm ion}$, and $M_{\rm ej,neu}$, all normalized by the initial
cloud mass $\Mcl$. \autoref{f:hst}(b) plots the temporal evolution
of the volume filling factor of the ionized gas
$\overline{x}_{\rm i} = \int (n_{\rm H^{+}}/n_{\rm H}) dV/\int dV$,
the escape fractions of ionizing ($f_{\rm esc,i}$) and non-ionizing
($f_{\rm esc,n}$) radiation, and the fraction of ionizing photons
absorbed by dust ($f_{\rm dust,i}$). These escape fractions and dust
absorption fraction here are the instantaneous probabilities of escape
or absorption from a single ray-trace at a given time.

\begin{figure*}[!t]
  \epsscale{1.26}\plotone{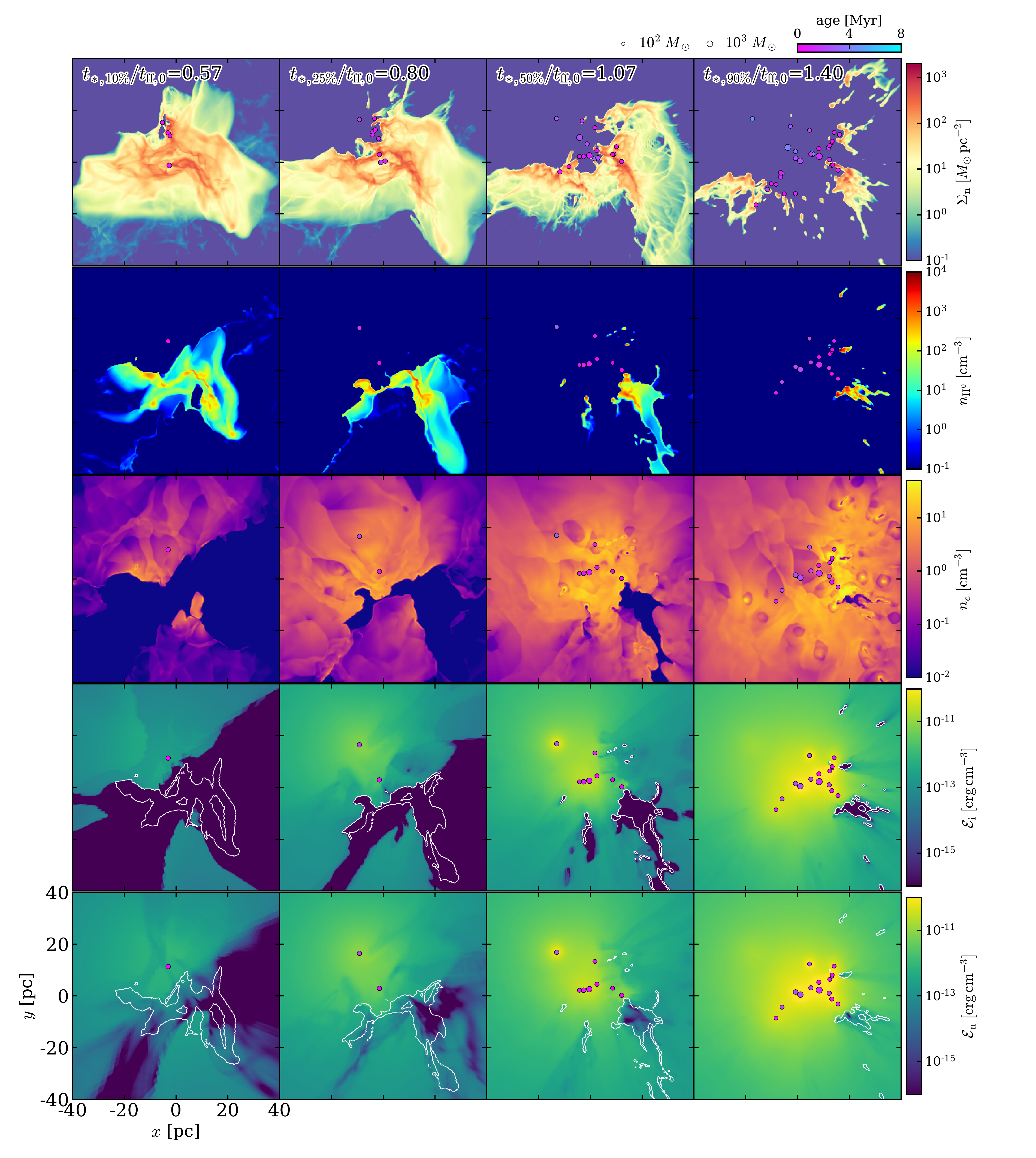}
  \caption{Snapshots of the fiducial model in the $x$-$y$ plane. The
    columns show times $t/\tffo=$0.58, 0.8, 1.07, and 1.40, and when
    10\%, 25\%, 50\%, 90\% of the final stellar mass has been
    assembled, respectively. From top to bottom: surface density of
    neutral gas (projected along the $z$-axis) $\Sigma_{\rm n}$,
    slices (passing through the position of the stellar center of
    mass) of neutral hydrogen number density $n_{\rm H^0}$, electron
    number density $n_{\rm e}$, energy density of ionizing radiation
    $\mathcal{E}_{\rm i}$, energy density of non-ionizing radiation
    $\mathcal{E}_{\rm n}$. The white contours in the bottom two rows
    show the number density of neutral gas at
    $n_{\rm H^0}=10\,{\rm cm}^{-3}$. Small circles in each frame mark
    the projected positions of the star particles that have formed,
    with their color and size corresponding to their age and mass,
    respectively. In the bottom four rows, only star particles whose
    distance from the slicing plane is less than $5\pc$ are
    shown.}\label{f:proj-slice}
\end{figure*}

\begin{figure*}
  \epsscale{1.2}\plotone{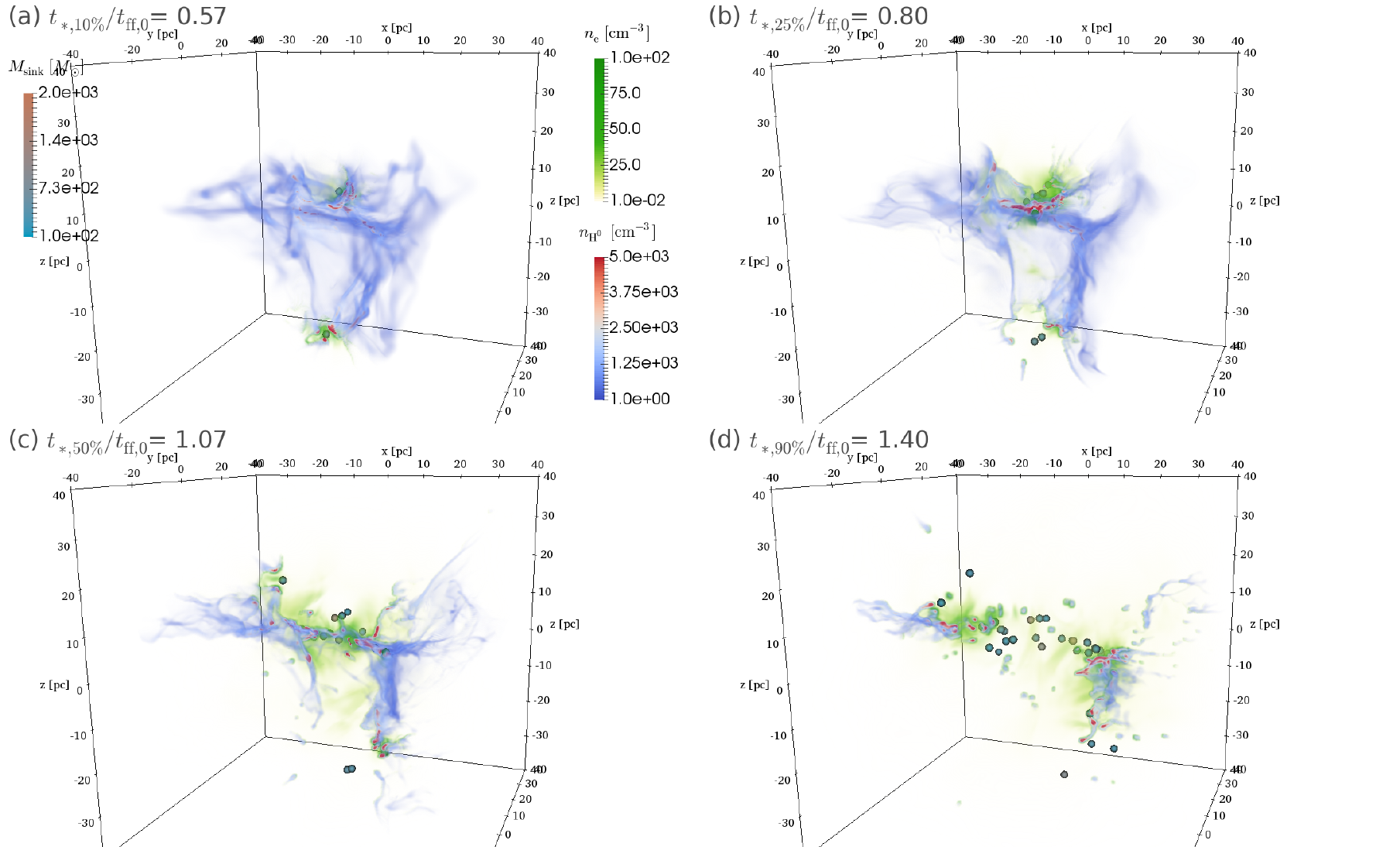}
  \caption{Volume rendering of the neutral hydrogen number density
    $n_{\rm H^0}$ (blue-white-red) and the electron number density
    $n_e$ (green-yellow) of the fiducial model at (a) $t=t_{*,10\%}$,
    (b) $t_{*,25\%}$, (c) $t_{*,50\%}$, and (d) $t_{*,90\%}$. Sink
    particles representing stellar clusters are shown as spheres
    colored by mass $M_{\rm sink}$.}\label{f:volrend}
\end{figure*}

\autoref{f:proj-slice} plots selected snapshots of the fiducial
model in the $x$-$y$ plane. From top to bottom, the rows show column
density of neutral gas projected along the $z$-axis, slices through
the stellar center of mass of the neutral hydrogen number density
$n_{\rm H^0}$, electron number density $n_{\rm e}$, ionizing radiation
energy density $\mathcal{E}_{\rm i}$, non-ionizing radiation energy
density $\mathcal{E}_{\rm n}$. The projected positions of star
particles are marked by circles colored by their ages. In the top row,
all star particles in the simulation volume are shown, while in the
bottom four rows only those within $\Delta 5 \pc$ of the slice are
shown. The selected times (columns from left to right) are $t/\tffo =$
0.57, 0.80, 1.07, and 1.40 when 10\%, 25\%, 50\%, and 90\% of the
final stellar mass has been assembled, respectively: here and
hereafter, these epochs will be referred to as $t_{*,10\%}$,
$t_{*,25\%}$, $t_{*,50\%}$, and $t_{*,90\%}$, respectively.
\autoref{f:volrend} displays volume rendered images of the fiducial
model at these epochs.

Because of the initial turbulence, the background medium is highly
clumpy and filamentary. Young \HII\ regions formed around star
particles quickly break out of their dense natal clumps, and in the
larger-scale turbulent cloud some lines of sight have quite low
optical depth. As a result, a substantial fraction of both ionizing
and non-ionizing photons escape from the simulation box even at quite
early stages of evolution. For example, at $t_{*,10\%}$ (only
$0.17\tffo = 0.8\Myr$ after the first collapse), five star particles
have been created and more than 50\% of computational box is filled
with ionized gas, although it accounts for only $3\%$ of the total gas
mass in the domain. By this time, 13\% of the radiation has escaped
from the domain overall (with an instantaneous escape probability of
16\%). Even though the second quadrant in the $x$-$y$ plane is rapidly
ionized, at early times (persisting through $t_{\rm *, 25\%}$) there
are still regions (see lower-left panels of \autoref{f:proj-slice})
that are fully neutral because they are shadowed by dense gas that
absorbs all the ionizing photons on rays emerging from the central
star-forming cluster.

At $t_{\rm *,25\%}$, star formation is concentrated in the central
$10 \pc$ region. Multiple \HII\ regions expand simultaneously and
merge with each other, accompanying a rapid increase of $\Mion$ by
intense photoevaporation. Due to thermal pressure gradients, the
ionized gas is pushed to the outer low-density regions, developing a
roughly exponential radial density profile outside the central cavity.
Ionized gas starts to escape from the box at mildly supersonic outflow
velocities of $\sim 20\kms$. Simultaneously, the remaining neutral gas
filaments are pushed away from the central collection of stars by a
combination of radiation pressure forces and the rocket effect as gas
photoevaporates from their surfaces (see top two rows of
\autoref{f:proj-slice}). At later time, star formation occurs
mainly in dense clumps within structures that have been pushed to the
periphery of the combined \HII\ region. We note that the expansion
over time of the loci of star formation is not primarily due to
shock-induced collapse as the \HII\ region expands (collapse in
filaments would have occurred anyway), but because dense gas is
progressively evacuated from a larger and larger volume.

\begin{figure*}[t]
  \epsscale{1.}\plotone{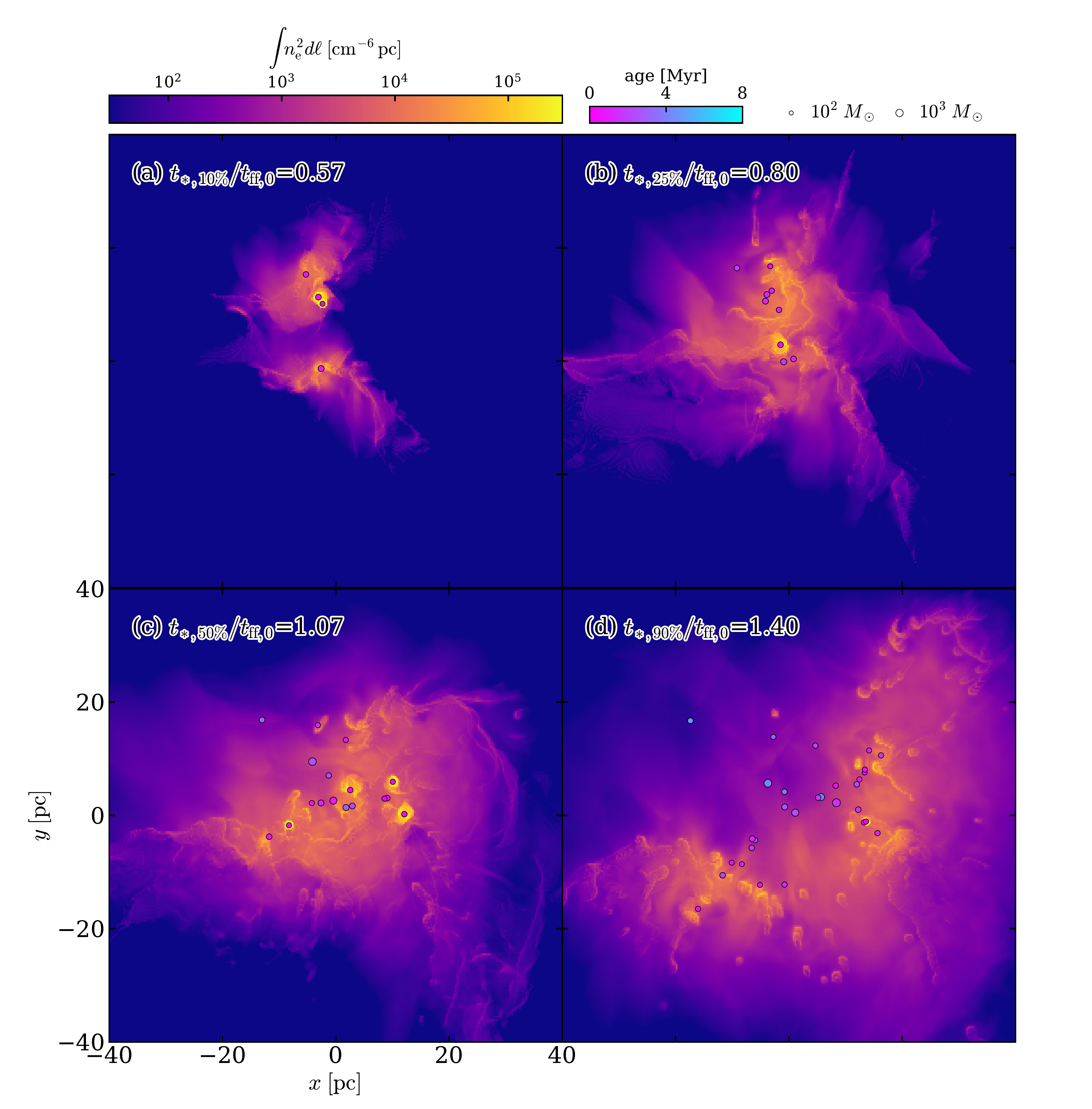}
  \caption{Snapshots projected on the $x$-$y$ plane of the emission
    measure of the fiducial model at (a) $t=t_{*,10\%}$, (b)
    $t_{*,25\%}$, (c) $t_{*,50\%}$, and (d) $t_{*,90\%}$. Circles mark
    the projected positions of all the star particles that have
    formed, with their color and size corresponding to their age and
    mass, respectively.}\label{f:nesq}
\end{figure*}

\autoref{f:nesq} plots the maps of the emission measure (EM)
$\int n_e^2 d\ell$ integrated along the $z$-direction of the fiducial
model at $t_{*,10\%}$, $t_{*,25\%}$, $t_{*,50\%}$, and $t_{*,90\%}$.
In the early phase of star formation, the EM map is dominated by
bright, compact \HII\ regions around newly formed stars, while at
later times it becomes rich in substructures resembling observed
pillars and bright-rimmed globules \citep[e.g.,][]{koe12,har15}; these
features are also apparent in the $n_e$ slices of
\autoref{f:proj-slice}. Similar structures have also been seen in
previous numerical simulations of expanding \HII\ regions in a
turbulent medium \citep[e.g.,][]{mel06,gri09,gri10,wal12}. The
globules exhibit ionization-driven ablation flows and distinctive
elongated tails pointing away from the ionizing sources. Anisotropic
mass loss, preferentially from the sides of globules facing the star
particles, exerts a recoil force on them; the globules rocket away
from the stellar sources at a typical radial velocity of
$v_{\rm ej,neu} \sim 5 \kms$.

By $t_{*,90\%}$, not only has the \HII\ region engulfed the entire
cloud, but the ionized region has expanded to twice the original cloud
radius. Furthermore, the central $10 \pc$ has mostly been cleared of
dense gas. The escape fractions of ionizing and non-ionizing photons
keep increasing with time as neutral gas surrounding ionizing sources
is pushed away \citep[see also, e.g.,][]{rog13,ras17}.

The cloud has been completely destroyed by $t/\tffo \sim 2$, with the
fraction of the remaining neutral gas $<2\%$ of the initial cloud
mass. The final star formation efficiency is
$\SFE\equiv M_{\rm *,final}/\Mcl=0.13$ in the fiducial model. Only 7\%
of the initial cloud mass is ejected in the neutral phase, while
$81\%$ is lost to photoevaporation. The remaining cluster of star
particles is loosely bound gravitationally and dissolves rapidly; 30\%
of the final stellar mass leaves the computational domain from
$t/\tffo=2.09$ to $t/\tffo=4$.

\subsection{Star Formation and Mass Loss Efficiencies}\label{s:eps}

\begin{figure*}
  \epsscale{1.0}\plotone{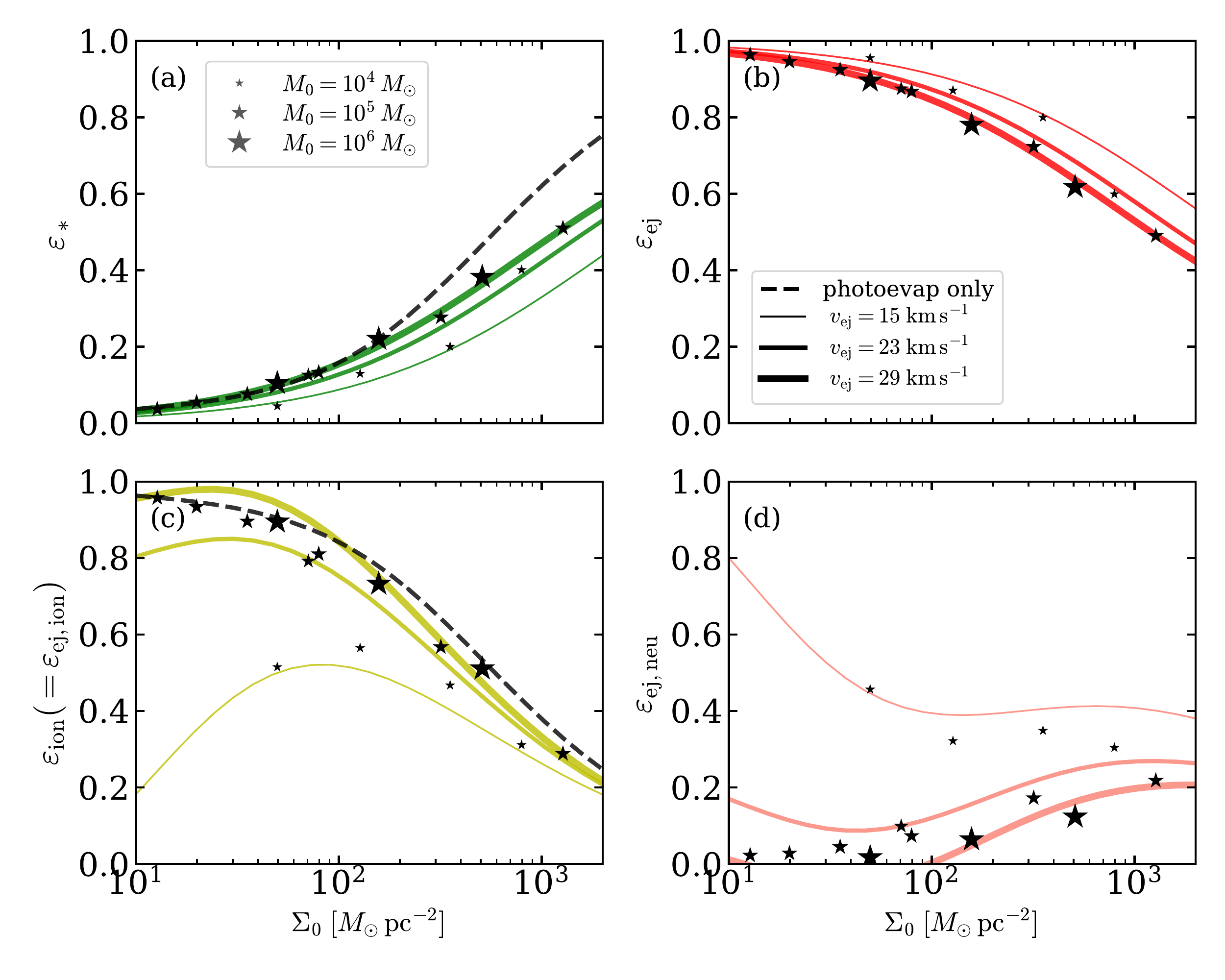}
  \caption{(a) Net star formation efficiency $\SFE$, (b) ejection
    efficiency $\epsej$, (c) photoevaporation efficiency $\epsion$,
    and (d) neutral ejection efficiency $\epsejneu$ for the PH+RP runs
    (star symbols) as functions of the initial gas surface density
    $\Sigmacl$. The symbol size is proportional to the initial cloud
    mass $\Mcl$ (this notation will be kept throughout this paper).
    Lines in each panel compare to our semi-analytic models for cloud
    disruption, as presented in Section~\ref{s:dispersal}: the dashed
    lines correspond to the prediction when photoevaporation dominates
    mass loss, while the solid lines are based on our numerical
    measurement of momentum injection applied to different cloud
    masses.}\label{f:eps-all}
\end{figure*}

\begin{figure}
  \epsscale{1.2}\plotone{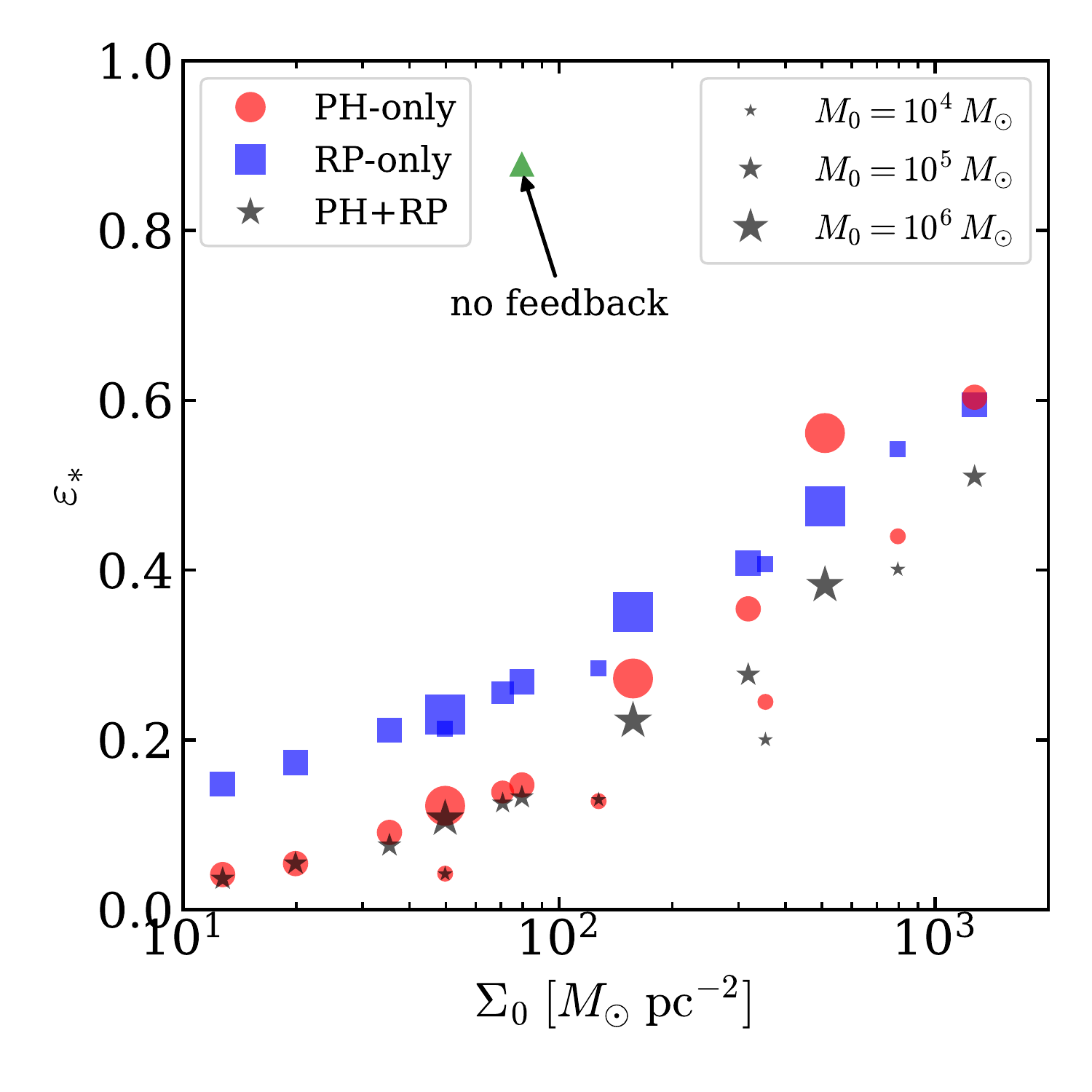}
  \caption{Net SFE $\SFE$ as a function of initial surface density
    $\Sigmacl$ for the models with photoionization only (PH-only, red
    circles), models with radiation-pressure only (RP-only, blue
    squares), and models with both photoionization and radiation
    pressure included (PH+RP, black stars). Symbol sizes denote the
    initial cloud mass. The green triangle represents $\SFE$ of the
    fiducial cloud with no feedback. For low surface density clouds,
    the net SFEs of PH-only runs are smaller than that of RP-only runs
    and are closer to that of PH+RP runs. This suggests that
    photoionization is more important than radiation pressure in cloud
    destruction. For dense and massive clouds ({\tt M1E5R05} and {\tt
      M1E6R25}), radiation pressure has a greater impact on cloud
    destruction.}\label{f:SFE-all}
\end{figure}

In our simulations, the SFE of a cloud rises until the associated
feedback photoevaporates and dynamically ejects the remaining gas,
completely halting further star formation. We define the final net SFE
as $\SFE\equiv M_{\rm *,final}/\Mcl$. In addition to the net SFE, we
similarly define photoevaporation and ejection efficiencies as
$\epsion \equiv M_{\rm ion,final}/\Mcl$ and
$\epsej \equiv M_{\rm ej,final}/\Mcl$, respectively, the latter of
which counts both neutral and ionized gases. Note that these
quantities are related by mass conservation through
$\epsej = 1 - \SFE = \epsion + \epsejneu$, where
$\epsejneu \equiv M_{\rm ej,neu,final}/\Mcl$ and $\epsion = \epsejion$
since no ionized gas is left inside the box at the end of the runs.

\autoref{f:eps-all} plots as star symbols (a) the net SFE $\SFE$,
(b) the ejection efficiency $\epsej$, (c) the photoevaporation
efficiency $\epsion$, and (d) the neutral ejection efficiency
$\epsejneu$ for all of our PH+RP runs as functions of the initial
surface density; $\SFE$, $\epsion$, and $\epsejneu$ are also tabulated
in Columns 2--4 of \autoref{t:result}. The size of the symbols
represents the initial cloud mass, a convention that will be adopted
throughout this paper. The net SFE depends weakly on $\Mcl$ but
strongly on $\Sigmacl$, increasing from $0.04$ for {\tt M1E5R50} to
$0.51$ for {\tt M1E5R05}. For clouds with low $\Sigmacl$ and
intermediate-to-high $\Mcl$ ({\tt M5E4R15}, {\tt M1E5R20}, {\tt
  M1E5R30}, {\tt M1E5R40}, {\tt M1E5R50}, {\tt M1E6R45}, {\tt
  M1E6R80}), photoevaporation is effective in destroying clouds:
$\epsion \gtrsim 0.7$ and the resulting SFE is $\SFE \lesssim 0.2$. By
contrast, the low mass clouds with $\Mcl = 10^4\Msun$ ({\tt M1E4R02},
{\tt M1E4R03}, {\tt M1E4R05}, {\tt M1E4R08}) and/or high surface
density ({\tt M1E5R10}, {\tt M1E5R05}, {\tt M1E6R25}) clouds have
$\epsejneu \gtrsim 0.1$, implying that dynamical ejection of neutral
gas is non-negligible for disruption and quenching of further star
formation in these clouds.

The effect of radiation feedback on cloud disruption is limited in the
highest-$\Sigmacl$ clouds such as {\tt M1E4R02}, {\tt M1E5R05} and
{\tt M1E6R25}. These reach $\SFE \gtrsim 0.4$ before the remaining gas
is dispersed. This is due to a deep gravitational potential well,
which outflowing gas must overcome to escape from the system
\citep[e.g.,][]{dal12}. These clouds also have a comparatively dense
and thick layer of recombining gas and dust that absorbs most ionizing
photons, resulting in relatively inefficient photoevaporative mass
loss (see Section~\ref{s:evap}). Thus, radiation pressure plays a
relatively more important role for the highest surface density clouds.

One can assess the relative (negative) impact of photoionization and
radiation pressure feedback on star formation by comparing the net
SFEs of the PH-only or RP-only runs with those of PH+RH models.
\autoref{f:SFE-all} plots $\SFE$ of PH-only (red circles), RP-only
(blue squares), and PH+RP (dark stars) models as a function of
$\Sigmacl$. Overall, the increasing trend of $\SFE$ with $\Sigmacl$ is
similar for all models. For most of our models, the net SFE of the
PH-only run is smaller than that of the RP-only counterpart and closer
to that of the PH+RP run, indicating that photoionization feedback
plays a more important role than radiation pressure in disrupting the
parent clouds. The two exceptions are the massive, high-surface
density clouds ({\tt M1E5R05} and {\tt M1E6R25}), for which the net
SFE of the RP-only run is smaller than that of the PH-only run.

The net SFEs of the RP-only runs lie approximately on a single
sequence as a function of $\Sigmacl$, regardless of $\Mcl$, also in
accordance with the recent theoretical predictions of the regulation
of star formation by the radiation pressure feedback alone
\citep{ras16,tho16}. We quantitatively show in Appendix~\ref{s:app2}
that our RP-only results are indeed in good agreement with the
analytic predictions. Our net SFE is slightly lower than those
obtained in the numerical simulations of \citet{ras16}, which employed
a grid-based moment method for radiation. \citetalias{kim17} showed
that this method does not adequately resolve the radiation field near
individual star particles (because radiation sources have finite
size), leading to more gas accretion and thus a higher SFE.

\autoref{f:SFE-all} also shows that model {\tt M1E5R20\_nofb}
(green triangle) in the absence of feedback converts $87\%$ of the gas
into stars over the course of evolution. The reason that even this
no-feedback run has a non-zero ejection efficiency of $\epsej = 0.13$
is that initial turbulence makes a small fraction of the gas
gravitationally unbound. \citet{ras16} found that the initial
``turbulent outflow ejection'' efficiency is
$\varepsilon_{\rm ej,turb}= 10$--$15\%$ for all models (see their
Figure 17a). In principle, this can be used to derive an ``adjusted''
feedback-induced SFE of
$\varepsilon_{\rm *,adj}= \SFE/(1-\varepsilon_{\rm ej,turb})$.

Finally, we note that the feedback effects of photoionizing radiation
and non-ionizing radiation on quenching star formation are not simply
additive. If feedback effects were additive, the naive expectation
would be
$1/\varepsilon_{\rm *,PH+RP}= 1/\varepsilon_{\rm *,PH} +
1/\varepsilon_{\rm *,RP}$. However, we find that the true value of
$\varepsilon_{\rm *, PH+RP}$ is $20$--$30\%$ larger than this naive
expectation.

\capstartfalse
\begin{deluxetable*}{cccccccccccc}\label{t:result}
\tabletypesize{\footnotesize}
\tablewidth{0pt}\tablecaption{Simulation Results}
\tablehead{
\colhead{Model} &
\colhead{$\SFE$} &
\colhead{$\epsion$} &
\colhead{$\epsejneu$} &
\colhead{$t_{*,0}$} &
\colhead{$t_{\rm SF}$} &
\colhead{$t_{\rm dest}$} &
\colhead{$t_{\rm ion}$} &
\colhead{$v_{\rm ej,neu}$} &
\colhead{$v_{\rm ej,ion}$} &
\colhead{$\langle f_{\rm ion} \rangle$} &
\colhead{$\langle q \rangle$} \\
\colhead{(1)} &
\colhead{(2)} &
\colhead{(3)} &
\colhead{(4)} &
\colhead{(5)} &
\colhead{(6)} &
\colhead{(7)} &
\colhead{(8)} &
\colhead{(9)} &
\colhead{(10)} &
\colhead{(11)} &
\colhead{(12)}}
\startdata
{\tt M1E5R50} & $0.04$ & $0.95$ & $0.02$ & $0.52$ & $0.41$ & $0.56$ & $0.60$ & $ 8.3$ & $24.4$ & $0.19$ & $  61$ \\
{\tt M1E5R40} & $0.05$ & $0.93$ & $0.03$ & $0.46$ & $1.03$ & $0.78$ & $0.75$ & $ 9.4$ & $24.3$ & $0.22$ & $  81$ \\
{\tt M1E5R30} & $0.08$ & $0.89$ & $0.04$ & $0.44$ & $0.72$ & $0.95$ & $0.98$ & $ 7.8$ & $23.9$ & $0.24$ & $ 116$ \\
{\tt M1E4R08} & $0.04$ & $0.52$ & $0.45$ & $0.56$ & $1.24$ & $1.04$ & $0.96$ & $ 6.3$ & $18.1$ & $0.42$ & $  69$ \\
{\tt M1E6R80} & $0.10$ & $0.89$ & $0.02$ & $0.33$ & $0.62$ & $0.69$ & $0.74$ & $10.7$ & $26.3$ & $0.37$ & $ 320$ \\
{\tt M5E4R15} & $0.12$ & $0.79$ & $0.09$ & $0.44$ & $1.14$ & $1.45$ & $1.31$ & $ 7.1$ & $23.1$ & $0.26$ & $ 148$ \\
{\tt \textbf{M1E5R20}} & $\bf 0.13$ & $\bf 0.81$ & $\bf 0.07$ & $\bf 0.40$ & $\bf 1.15$ & $\bf 1.39$ & $\bf 1.32$ & $\bf  7.9$ & $\bf 24.1$ & $\bf 0.26$ & $\bf  203$ \\
{\tt M1E4R05} & $0.13$ & $0.56$ & $0.32$ & $0.52$ & $1.19$ & $2.07$ & $1.61$ & $ 6.8$ & $20.2$ & $0.36$ & $ 199$ \\
{\tt M1E6R45} & $0.22$ & $0.73$ & $0.06$ & $0.27$ & $1.12$ & $1.37$ & $1.44$ & $13.2$ & $28.6$ & $0.37$ & $ 753$ \\
{\tt M1E5R10} & $0.28$ & $0.56$ & $0.17$ & $0.33$ & $1.79$ & $2.26$ & $2.08$ & $ 8.3$ & $25.6$ & $0.42$ & $ 807$ \\
{\tt M1E4R03} & $0.20$ & $0.47$ & $0.34$ & $0.44$ & $2.13$ & $3.16$ & $2.48$ & $ 6.0$ & $20.9$ & $0.36$ & $ 402$ \\
{\tt M1E6R25} & $0.38$ & $0.52$ & $0.11$ & $0.24$ & $1.70$ & $2.41$ & $2.38$ & $14.7$ & $35.6$ & $0.43$ & $2032$ \\
{\tt M1E4R02} & $0.40$ & $0.32$ & $0.29$ & $0.40$ & $2.32$ & $4.10$ & $2.97$ & $ 6.7$ & $24.1$ & $0.46$ & $1630$ \\
{\tt M1E5R05} & $0.51$ & $0.31$ & $0.19$ & $0.27$ & $2.35$ & $3.59$ & $3.46$ & $ 9.7$ & $30.2$ & $0.50$ & $2950$ \\
\tableline
{\tt M1E5R20\_N128} & $0.15$ & $0.79$ & $0.07$ & $0.42$ & $0.92$ & $1.13$ & $1.26$ & $ 8.7$ & $23.3$ & $0.25$ & $ 261$ \\
{\tt M1E5R20\_N512} & $0.12$ & $0.82$ & $0.07$ & $0.38$ & $1.35$ & $1.53$ & $1.39$ & $ 7.6$ & $24.2$ & $0.26$ & $ 188$ \\
{\tt M1E5R20\_nofb} & $0.88$& \nodata & $0.13$ & $0.40$ & $5.85$& \nodata& \nodata & $ 3.9$& \nodata& \nodata& \nodata 
\enddata
\tablecomments{Column 1: model name. Column 2: net star formation
  efficiency. Column 3: photoevaporation efficiency. Column 4: neutral
  ejection efficiency. Column 5: time of first star formation (in
  units of $\tffo$). Column 6: star formation duration
  $t_{\rm SF} = t_{*,95\%} - t_{*,0}$ (units $\tffo$). Column 7:
  timescale for cloud destruction
  $t_{\rm dest} = t_{\rm neu,5\%} - t_{*,0}$ (units $\tffo$). Column
  8: timescale for photoevaporation (units $\tffo$;
  Equation~\eqref{e:tion}). Column 9: time-averaged outflow velocity
  of the neutral gas ($\rm km\,s^{-1}$). Column 10: time-averaged
  outflow velocity of the ionized gas ($\rm km\,s^{-1}$). Column 11:
  time-averaged hydrogen absorption fraction (see
  Equation~\eqref{e:fionavg}). Column 12: time-averaged shielding
  factor. The fiducial model {\tt M1E5R20} is shown in
  bold.}\label{t:result}
\end{deluxetable*}
\capstarttrue

\begin{figure*}[!t]
  \epsscale{1.0}\plotone{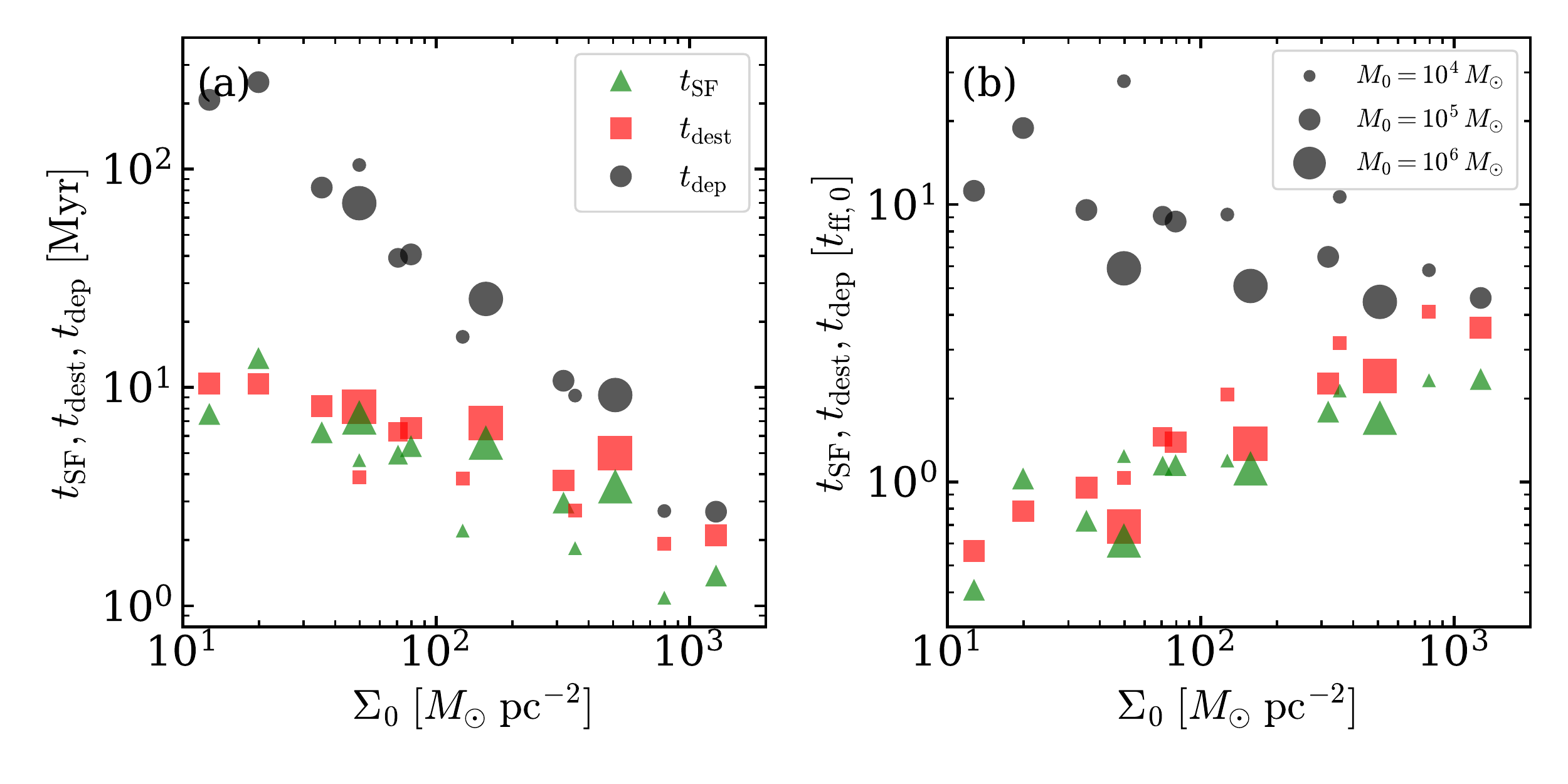}
  \caption{(a) Timescales for star formation
    $t_{\rm SF} = t_{*,95\%} - t_{*,0}$ (triangles), cloud destruction
    $t_{\rm dest} = t_{\rm neu,5\%} - t_{*,0}$ (squares), and
    effective gas depletion $t_{\rm dep} = t_{\rm SF}/\SFE$ (circles)
    for all PH+RP models as functions of the initial cloud surface
    density $\Sigmacl$. (b) Same as (a) but with the timescales
    normalized by the initial freefall time
    $\tffo$.}\label{f:timescales}
\end{figure*}

\subsection{Timescales for Star Formation and Cloud Destruction}

\begin{figure*}
  \epsscale{1.0}\plotone{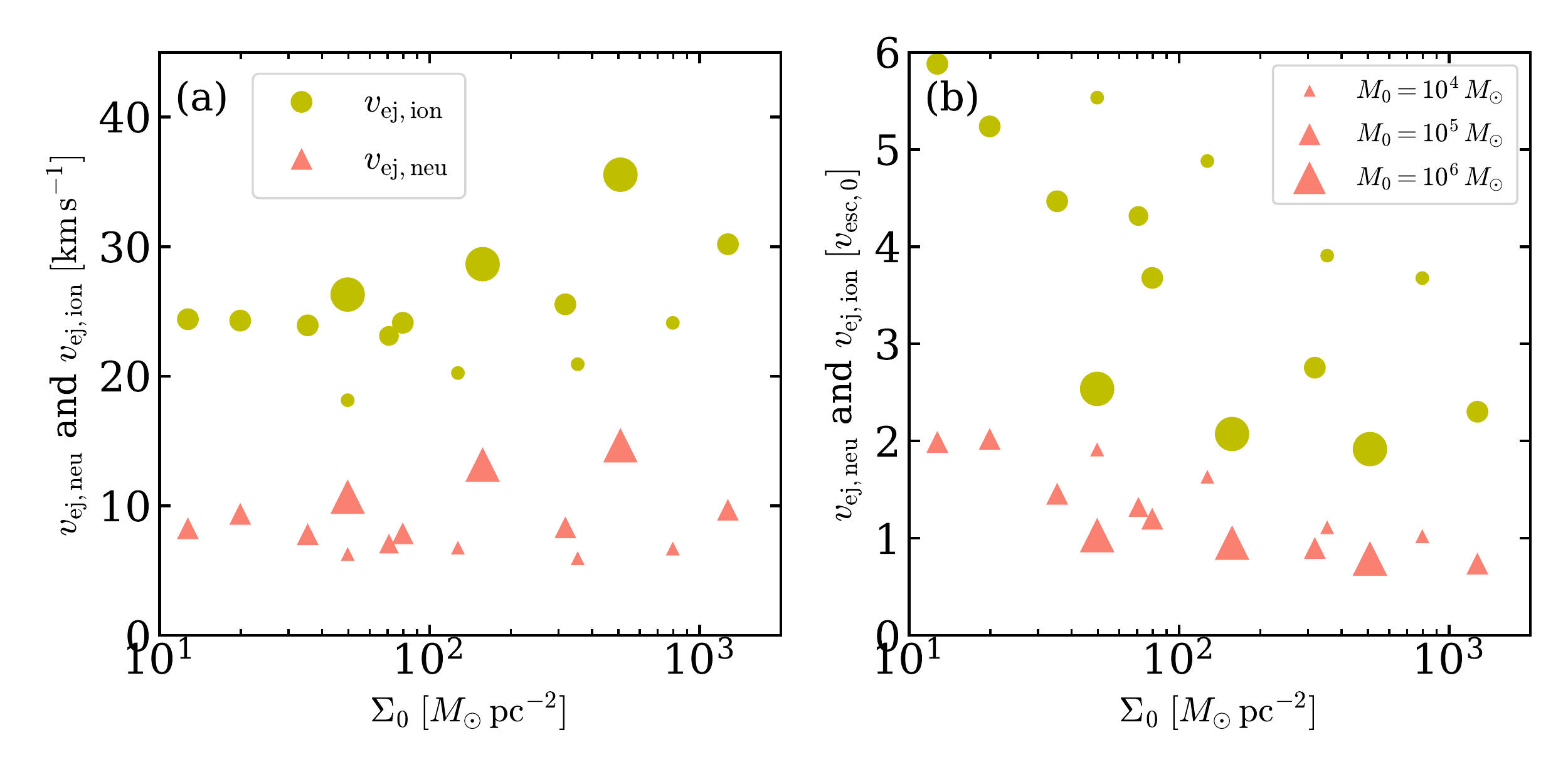}
  \caption{(a) Mean radial velocity of the outflowing neutral gas
    (triangles) and ionized gas (circles) for all PH+RP models as
    functions of the initial cloud surface density $\Sigmacl$. (b)
    Same as (a) but normalized by the initial escape velocity
    $\vesc$.}\label{f:vej}
\end{figure*}

The cloud lifetime is important to understanding the life cycle of
GMCs within the context of other ISM phases, yet observational
constraints on it remain quite uncertain \citep{hey15}. We use our
simulations to directly measure the relevant timescales as follows. We
define the star formation timescale in our models as the time taken to
assemble 95\% of the final stellar mass after the onset of radiation
feedback, i.e., $t_{\rm SF} \equiv t_{*,95\%} - t_{*,0}$. The
approximate gas depletion timescale by star formation is then
$t_{\rm dep} = t_{\rm SF}/\SFE$. We regard a cloud as being
effectively destroyed at $t=t_{{\rm neu,5\%}}$ when only 5\% of the
initial cloud mass is left over as the neutral phase in the simulation
box, with the other 95\% transformed into stars, photoionized, or
ejected. We define the destruction timescale as
$t_{\rm dest} = t_{{\rm neu},5\%} - t_{*,0}$.\footnote{For the {\tt
    M1E4R08} model, we take $t_{\rm dest}$ as the time difference
  between $t_{{\rm neu},5\%}$ and the time at which second sink
  particle formed, because the first star particle, with its small
  ionizing photon output rate
  $\Qi \sim 2.1 \times 10^{46}\second^{-1}$, has limited impact on the
  cloud evolution (see also \autoref{f:hst-dot-Mevap}).} In most
cases, $t_{\rm dest} > t_{\rm SF}$,\footnote{In the {\tt M1E5R40}
  model, late-time star formation (roughly $\sim 10\%$ of the final
  stellar mass) occurs in a dense globule even after
  $t_{\rm neu,5\%}$.} but generally the two timescales are very
similar. Columns 5--7 of \autoref{t:result} list $t_{*,0}$,
$t_{\rm SF}$, and $t_{\rm dest}$, respectively.

\autoref{f:timescales} plots $t_{\rm SF}$ (triangles),
$t_{\rm dest}$ (squares), and $t_{\rm dep}$ (circles) of our PH+RP
runs in units of Myr (left panel) and in units of the initial freefall
time (right panel). In physical units, $t_{\rm SF}$ and $t_{\rm dest}$
increase from $\sim 1$--$3\Myr$ in the densest clouds
($n_{\rm H,0} \gtrsim 500 \;{\rm cm}^{-3}$ and
$\Sigmacl \gtrsim 300 \Sunit$) to $\sim 8$--$14 \Myr$ in the
lowest-density clouds ($n_{\rm H,0} \lesssim 10 \;{\rm cm}^{-3}$ and
$\Sigmacl \lesssim 20 \Sunit$). Relative to the freefall time,
$t_{\rm SF}$ ($t_{\rm dest}$) decreases from $\sim 2\tffo$
($\sim 4\tffo$) in the dense clouds to $\sim 0.4 \tffo$
($\sim 0.6 \tffo$) in the low-density clouds. Our result that
$t_{\rm SF} \lesssim 2\tffo$ is consistent with the picture of rapid
star formation envisaged by \citet{elm00}. The depletion timescale
decreases considerably with $\Sigmacl$ from $250 \Myr$ for
low-$\Sigma_0$ model ({\tt M1E5R40}) to $2.7 \Myr$ for high-$\Sigma_0$
models ({\tt M1E4R02} and {\tt M1E5R05}).

\subsection{Outflow Velocity}\label{s:vej}

In the presence of ionizing radiation, the outflows in our simulations
consist of both neutral and ionized gas, and it is interesting to
measure the dependence of ejection velocities on cloud properties. We
calculate the mass-weighted ejection velocities of neutral/ionized gas
as
\begin{equation}
  v_{{\rm ej,neu/ion}}
  = \dfrac{p_{{\rm ej,neu/ion,final}}}{M_{{\rm ej,neu/ion,final}}}\,,
\end{equation}
where
\begin{equation}
  p_{{\rm ej,neu/ion,final}}
  = \int dt \int_{\partial V} dA\, \rho_{\rm neu/ion} (\mathbf{v}
  \cdot \mathbf{\hat{n}}) (\mathbf{v} \cdot \mathbf{\hat{r}})
\end{equation}
is the time-integrated total radial momentum of neutral/ionized
outflowing gas with $\mathbf{\hat{r}}$ being the unit radial vector
with respect to the stellar center of mass and $\mathbf{\hat{n}}$
being the unit vector normal to the surface area $dA$ at the outer
boundary of the box. These values are given in Columns 9 and 10 of
\autoref{t:result}.

\autoref{f:vej} plots $v_{\rm ej,neu}$ (triangles) and
$v_{\rm ej,ion}$ (circles) in units of ${\rm km\,s}^{-1}$ (left panel)
and in units of the initial escape velocity $\vesc$ (right panel) for
all the PH+RP runs. The outflow velocity of the ionized gas is
$\sim 18$--$36\kms$ and larger than $\vesc$ across the whole range of
$\Sigmacl$. In low-surface density clouds, these supersonic outflows
of the ionized gas are driven primarily by thermal pressure. Since
ionized gas has low optical depth and hence high Eddington ratios, its
ejection in high-surface density clouds is further helped by radiation
pressure, while strong gravity tends to reduce the outflow velocity.

The neutral gas is ejected at a typical velocity of $\sim
6$--$15 \kms$, about $\sim 2$--$4$ times smaller than the ejection
velocity of the ionized gas. This is roughly consistent with the
characteristic rocket velocity $\sim 5$--$10\kms$ of cometary globules
in a photoevaporation-dominated medium \citep[e.g.,][]{ber90},
although the outflow velocity is also affected by gravity as well as
radiation pressure forces especially for high-surface density clouds.
Note that $v_{\rm ej,neu}/\vesc \sim 0.7$--$2.0$ decreases slowly with
increasing $\Sigmacl$ owing to gravity.

\section{MASS LOSS PROCESSES}\label{s:ML}

For all of our simulations, radiative stellar feedback leads to at
least half of the original mass in the cloud being ejected. Thus,
radiation from newly formed stars actively quenches future star
formation. As \autoref{f:eps-all} shows, most of the ejected gas is
in the ionized phase, especially for typical GMCs in normal disk
galaxies with $\Sigmacl \lesssim 10^2 \Sunit$ and
$\Mcl \gtrsim 10^5 \Msun$. For high-density or low-mass clouds,
ejection of the neutral phase is non-negligible, but $\epsejneu$ never
exceeds $\epsejion$. In this section, we first develop and calibrate a
simple theoretical models of mass loss by photoevaporation. We then
analyze the momentum injection in our models, including both ionized
and neutral gas. These results for photoevaporation and momentum
injection will be used in Section \ref{s:dispersal} to make
predictions for the net SFEs and mass-loss efficiencies in comparison
with our numerical results.

\subsection{Mass Loss by Photoevaporation}\label{s:evap}

\begin{figure*}
  \epsscale{0.8}\plotone{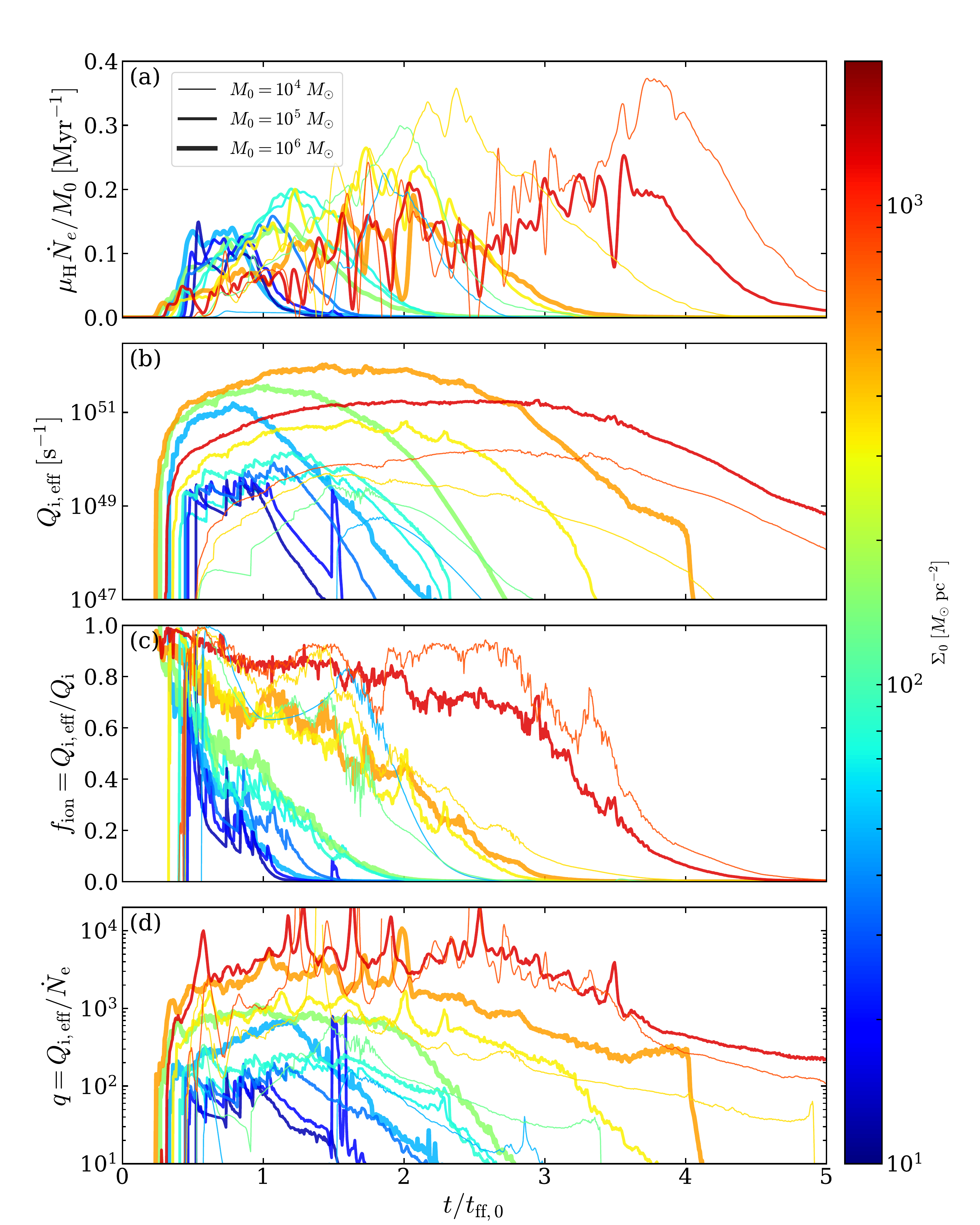}
  \caption{(a) Mass loss rate $\muH\dotNe$ via photoevaporation
    normalized by the initial cloud mass $\Mcl$, (b) total
    photoionization rate $\Qieff$, (c) fraction of ionizing photons
    absorbed by neutral hydrogen $f_{\rm ion} = \Qieff/\Qi$, and (d)
    shielding factor $q=\Qieff/\dotNe$ as functions of time. The line
    color and thickness respectively correspond to the surface density
    and mass of the initial cloud. The shielding factor is much larger
    than unity in all models, indicating that most ionizing photons
    are used in maintaining ionization equilibrium within the \HII\
    region rather than creating new ionized
    gas.}\label{f:hst-dot-Mevap}
\end{figure*}

We first consider the case in which star formation is quenched solely
by photoevaporation. Ionizing photons emitted by star particles arrive
at ionization fronts (IFs) after passing through an intervening volume
in which the gas is nearly fully ionized, with a balance between
ionization and recombination. Ionized gas created at IFs does not
participate in star formation (provided it remains sufficiently
exposed to radiation) and will be eventually pushed out of the cloud
and simulation volume by thermal and radiation pressures. Here we use
physical scaling arguments to estimate the net photoevaporation rate
in a star-forming cloud.

\subsubsection{Ionization Analysis}

\HII\ regions formed in our simulations are highly irregular in shape.
The EM map is dominated by pillars, ridges, and cometary globules near
the \HII\ region peripheries, as \autoref{f:nesq} illustrates. Each
of these bright features marks the IF on the surface of an
optically-thick, neutral gas structure that is facing the ionizing
sources. These neutral structures are photoevaporated as the IF
advances. Since the newly ionized gas can expand into the
surroundings, the IFs are typically D-critical or of strong-D type,
and the ionized gas streams off the IF at approximately the sound
speed and forms an ionized boundary layer (e.g.,
\citealt{oor55,kah69,elm76,ber89}). In addition to the bright IF
regions, Figures~\ref{f:proj-slice}-\ref{f:nesq} show that there is
substantial diffuse ionized gas in the volume, filling the space
between and beyond the filaments and clumps of neutral gas. Along some
lines of sight from stellar sources, ionizing photons moving toward
optically-thin regions produce a weak-R type IF that quickly
propagates out of the simulation domain, such that in those directions
the \HII\ region is effectively density bounded. In other directions,
lines of sight from ionizing sources propagate through diffuse ionized
gas until they end at an IF on the surface of dense neutral structure.
As we will see below, almost all of the ionizations and recombinations
in our simulations occur in the ionization-bounded Str\"{o}mgren
volume (IBSV), and ionization and recombination are in global
equilibrium.

The equation for net electron production integrated over the entire
domain reads
\begin{equation}\label{e:cont}
  \dotNe = \int \mathcal{I}\,dV - \int \mathcal{R}\,dV\,,
\end{equation}
where
$\mathcal{I} = n_{\rm H^0}(c\mathcal{E}_{\rm i}/h\nu_{\rm i})
\sigma_{\rm H^0}$ is the photoionization rate per unit volume with $c$
being the speed of light;
$\mathcal{R} = \alphaB n_{\rm e}n_{\rm H^{+}}$ is the recombination
rate per unit volume, with
$\alphaB = 3.03 \times 10^{-13} (T/8000\Kel)^{-0.7}
\cm^{3}\second^{-1}$ being the case B recombination coefficient. The
rate at which ionized gas is newly created within the domain (from
ionizations exceeding recombinations) is $\dotMion = \mu_H \dotNe$,
where $\muH=1.4 m_{\rm H}$ is the mean atomic mass per hydrogen. Part
of this may go into an increase in the mass of ionized gas within the
box, and part may go into escape of ionized gas from the domain:
$\dotMion = \dot{M}_{\rm ion,box} + \dot{M}_{\rm ion,ej}$.

In our simulations, some fraction of the ionizing photons emitted by
the stars are absorbed by neutral hydrogen, while others are absorbed
by dust or escape from the simulation domain. Let
$f_{\rm ion} = 1 - f_{\rm esc,i} - f_{\rm dust,i}$ denote the fraction
of ionizing photons absorbed by neutral hydrogen. Then, the total
photoionization rate of neutral hydrogen inside the \HII\ region can
be written as
\begin{equation}\label{e:Qieff}
  \int \mathcal{I}\,dV  
  = f_{\rm ion}\Qi \equiv \Qieff\,.
\end{equation}

Newly-ionized gas is created at IFs, and we denote the total effective
area of these IFs as $\Ai$. Gas streams away from the IFs with typical
number density $\nion$ and characteristic velocity
$\ci= (2.1 \kB T_{\rm ion}/\muH)^{1/2} = 10.0 \kms$
\citep[e.g.,][]{ber89}. We thus write
\begin{equation}\label{e:dotNe}
\dotNe \equiv  \ci\nion\Ai\,
\end{equation}
for the rate of election production at the IFs.
Equation~\eqref{e:dotNe} may be thought of as defining the product
$\nion\Ai$ in terms of $\dotNe$.

To reach the IFs, photons traverse a volume of nearly fully ionized
gas that is undergoing recombination (which is approximately balanced
by ionization; see below). If we define the effective path length
through this ionized volume as $\Hi$, we can write the total
recombination rate as
\begin{equation}\label{e:rec}
  \int\mathcal{R}\,dV  \equiv  \alphaB\nion^2 \Ai\Hi\,.
\end{equation}
With $\nion\Ai$ defined via Equation~\eqref{e:dotNe}, we can think of
Equation~\eqref{e:rec} as defining the ratio $\Ai/\Hi$, which has
units of length.

The volume of highly-ionized gas between the source and the IF acts as
an insulator, absorbing photons that would otherwise reach the IF in
order to balance recombination. We define the ratio of available
ionizing photons to photons that actually reach the IF as a shielding
factor:
\begin{equation}
  q \equiv \dfrac{\Qieff}{\dotNe} = 1 + \frac{\int
    \mathcal{R}\,dV}{\int \mathcal{I}\,dV - \int \mathcal{R}\,dV} =
  1 + \dfrac{\alphaB\nion\Hi}{\ci}\label{e:q2} \,.
\end{equation}
With this definition, $q=1$ would imply all the ionizing photons are
used to ionize neutrals at the IF, while $q \gg 1$ when most of the
ionizing photons are shielded by the IBSV. The second term on the
right-hand side of Equation~\eqref{e:q2}, $\alphaB\nion\Hi/\ci$,
represents the ratio of the characteristic flow timescale to the
recombination timescale, or the number of recombinations occurring
over the time it would take the flow to cross the IBSV.

\autoref{f:hst-dot-Mevap} plots the temporal histories of (a) the
specific evaporation rate $\muH\dotNe/\Mcl$, (b) the total
photoionization rate $\Qieff$, (c) hydrogen absorption fraction of
ionizing photons $f_{\rm ion}$, and (d) the shielding factor $q$, for
all of our PH+RP runs. Both the evaporation rate and photoionization
rate reach peak values and then decline with time as the cloud is
destroyed.

The hydrogen absorption fraction is largest in the early, embedded
phase of star formation and decreases with time as optical depth drops
and most of the ionizing photons escape the computational box. Column
(11) of \autoref{t:result} gives
\begin{equation}\label{e:fionavg}
  \langle f_{\rm ion} \rangle \equiv \frac{\int \dotNe f_{\rm ion}\,dt
  }{\int \dotNe\,dt}\,,
\end{equation}
where the angle brackets $\langle\,\rangle$ denote the
$\dotNe$-weighted temporal average over the whole simulation period
after $t_{*,0}$ (when $\dotNe$ is nonzero). The value of
$\langle f_{\rm ion} \rangle$ ranges from $0.19$ to $0.50$. Overall, a
larger fraction of ionizing radiation is absorbed by hydrogen in
higher surface density clouds as \HII\ regions undergo a relatively
longer embedded phase.

Figure \ref{f:hst-dot-Mevap} shows that the shielding factor is much
larger than unity for all models at all times. Column (12) of
\autoref{t:result} gives $\dotNe$-weighted time-averaged values
$\langle q \rangle$, which ranges from $61$ to $2950$, increasing with
mass and surface density. This is consistent with the expectation (cf.
the right-hand side of Equation~\eqref{e:q2}) that a higher
recombination rate in denser clouds, as well as longer path lengths in
larger clouds, increases shielding and therefore reduces the
efficiency of photoevaporation.

The fact that our clouds have $q \gg 1$ justifies the assumption of a
global equilibrium between ionization and recombination
\begin{equation}\label{e:IBL1}
  \Qieff=\int \mathcal{I}\,dV \approx \int \mathcal{R}\,dV
      = \alphaB\nion^2 \Ai \Hi \,.
\end{equation}
We solve this to obtain
$\nion \approx [\Qieff/(\alphaB \Ai\Hi)]^{1/2}$, and from
Equation~\eqref{e:q2} we obtain the shielding factor
\begin{equation}
  q \approx \dfrac{\alphaB\nion\Hi}{\ci}
  \approx \dfrac{\alphaB^{1/2}}{\ci}
  \left(\dfrac{\Qieff\Hi}{\Ai}\right)^{1/2}\,,
\end{equation}
while Equation~\eqref{e:dotNe} yields an approximate expression for the
evaporation rate:
\begin{equation}\label{e:dotNe-approx}
  \dotNe \approx \dfrac{\ci}{\alphaB^{1/2}}
  \left(\dfrac{\Qieff\Ai}{\Hi} \right)^{1/2}\,.
\end{equation}

The timescale for the photoevaporative mass loss tends to be longer in
high surface density clouds, similarly to $t_{\rm dest}$ and
$t_{\rm SF}$. Despite the fact that they have a relatively high
hydrogen absorption fraction and a small escape fraction of ionizing
photons, the mass loss is inefficient because of large $q$. In
addition, the motions of sink particles and the associated neutral
envelopes cause a sudden change in the optical depth of ionizing
photons. As a result, the size and shape of the \HII\ region and
photoevaporation rate fluctuate with time until they completely
exhaust accretion flows and break out to larger radii at late time
\citep[e.g.,][]{pet10,dal12}.

We shall define the cloud photoevaporation timescale as
\begin{equation}\label{e:tion}
\tion \equiv \dfrac{\Ne}{\langle \dotNe \rangle} =
    \dfrac{\Ne^2}{\int\dotNe^2\,dt}\,,
\end{equation}
where $\Ne = \int \dotNe\,dt$ is the integrated number of electrons.
Column (8) of \autoref{t:result} lists the value of $\tion$ for all
models. Our numerical simulations show that this timescale is
typically comparable to the initial freefall timescale for the cloud,
$\tffo$.

\subsubsection{Constraints from Simulations}\label{s:phifits}

\begin{figure}
  \epsscale{1.1}\plotone{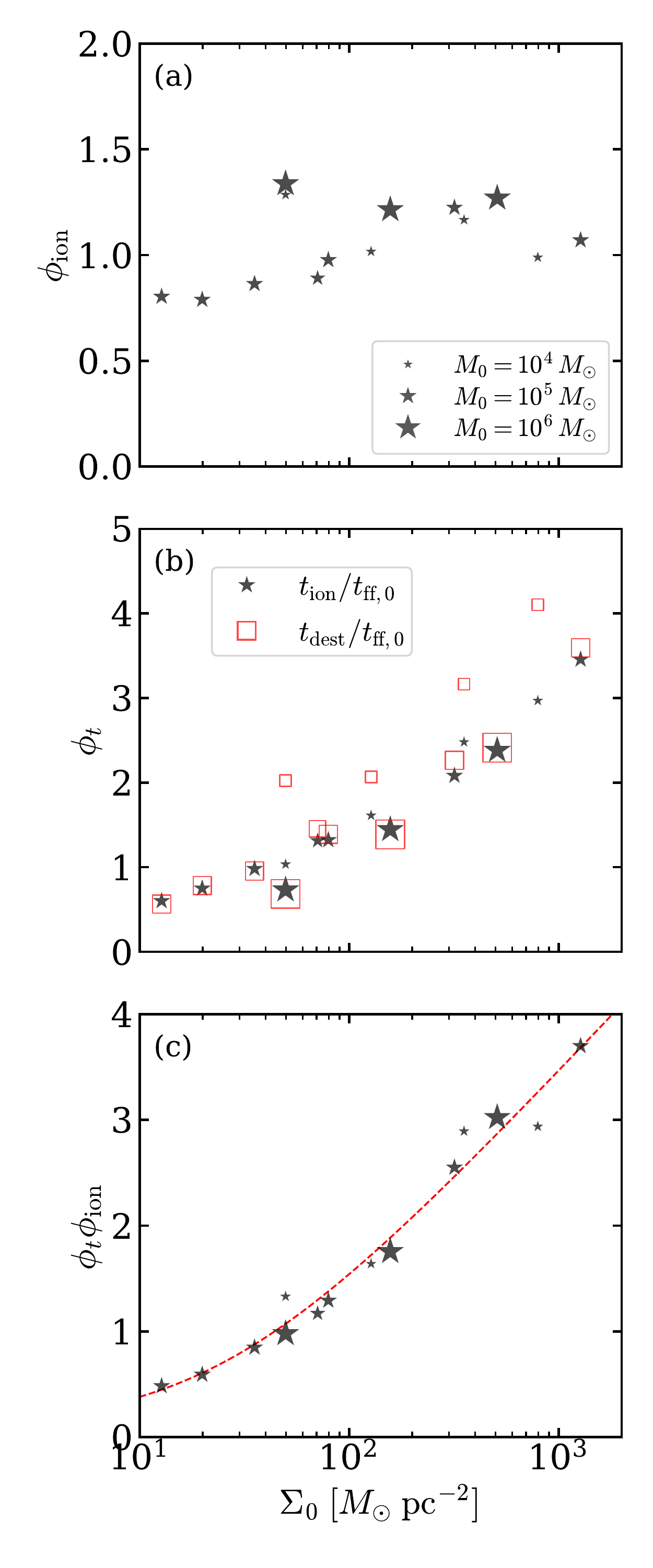}
  \caption{Dependence on $\Sigmacl$ of (a) the dimensionless
    ionization rate $\phiion$, (b) the dimensionless photoevaporation
    timescale $\phit$ for clouds, and (c) the product $\phiion\phit$,
    displayed as star symbols for all PH+RP models. In (b), the cloud
    destruction timescale $t_{\rm dest}/\tffo$ is shown for comparison
    as open squares. The red dashed line in (c) is our fit to the
    numerical results (see Equation~\ref{e:phifit}).}\label{f:phiion}
\end{figure}

Once $\Qieff$, $\Ai$, and $\Hi$ are known,
Equation~\eqref{e:dotNe-approx} can be used to estimate the
photoevaporation mass loss rate in star-forming clouds. It is natural
to expect that the total area $\Ai$ of the ionization front scales
with the surface area $4 \pi R_0^2$ of the cloud, while the thickness
$\Hi$ of the IBSV scales with the cloud radius $R_0$. Our numerical
results, based on measured $\Qieff$, $\dotNe$, and
$\int n_e^2 dV/\int n_e dV\rightarrow \nion$, show that time-averaged
values of $\Ai /(4\pi\Rcl^2) = \dotNe/(4\pi \nion\ci\Rcl^2)$ and
$ \Hi /\Rcl \approx \ci\Qieff/(\alphaB\nion\dotNe\Rcl)$ are indeed of
order unity. In detail, the time-averaged values of $\Ai/(4\pi\Rcl^2)$
and $\Hi/\Rcl$ and are in the ranges $0.5$--$2.0$ and $0.6$--$1.9$,
respectively. Motivated by Equation~\eqref{e:dotNe-approx}, together
with the characteristic dynamical time of clouds, we introduce the
dimensionless evaporation rate $\phiion$ and evaporation timescale
$\phit$ defined as
\begin{align}
  \phiion & \equiv \langle \dotNe \rangle\times \displaystyle
            \frac{\alphaB^{1/2}}{\ci Q_{\rm i,max}^{1/2}\Rcl^{1/2}}\,,
            \label{e:phiion} \\
  \phit & \equiv  \displaystyle \frac{\tion}{\tffo}
          \,, \label{e:phit}
\end{align}
where $Q_{\rm i,max}$ is the maximum rate of the total ionizing
photons over the lifetime of a star cluster (which is directly related
to the SFE).

\autoref{f:phiion} plots $\phiion$ and $\phit$ from our PH+RP
models as star symbols. Note that $\phiion \sim 0.79$--$1.34$, roughly
independent of the cloud mass and surface density. From the definition
in Equation~\eqref{e:phiion}, an order-unity value of $\phiion$
implies a mass-loss rate from clouds in which the characteristic
velocity is the ionized-gas sound speed, the characteristic spatial
scale is the size of the cloud, and the characteristic density is
consistent with ionization-recombination equilibrium; comparison to
Equation~\eqref{e:dotNe-approx} implies that
$\langle (\Qieff/Q_{\rm i,max}) (\Ai/\Hi)\rangle^{1/2}/\Rcl^{1/2}
\approx \phiion \sim 1$.

\autoref{f:phiion} shows that $\phit$ ranges between $0.6$ to $3.5$
and increases with $\Sigmacl$. A typical photoevaporation timescale is
thus twice the freefall time in the cloud. Plotted as open squares is
the normalized destruction timescale $t_{\rm dest}/\tffo$. This is
very close to $\phit$ for low-density, massive clouds, suggesting that
these clouds are destroyed primarily by photoevaporation. Low-mass
and/or high-surface density clouds have a destruction timescale
somewhat longer than the evaporation timescale, suggesting that cloud
destruction involves ejection of neutrals (accelerated by a
combination of rocket effect and radiation pressure) as well as ions.

The total number of electrons (and hence ions) created by
photoevaporation over the lifetime of the cloud is
$\Ne = \int \dotNe dt \equiv \langle \dotNe \rangle t_{\rm ion}$. The
total mass photoevaporated from the cloud can therefore be expressed
as
\begin{equation}\label{e:Mion}
  M_{\rm ion} = \muH  N_e = \phit\phiion
  \frac{\muH \ci Q_{\rm i,max}^{1/2}\Rcl^{1/2}\tffo}{\alphaB^{1/2}}.
\end{equation}
Since the right-hand side is proportional to
  $M_{*,{\rm final}}^{1/2}$, this shows that the fraction of mass
photoevaporated over the cloud lifetime, $M_{\rm ion}/\Mcl$, scales as
the square root (rather than linearly) of the star formation
efficiency. We discuss this result further in
Section~\ref{s:dest_evap}. \autoref{f:phiion}(c) plots the product
$\phit\phiion$ as a function of $\Sigmacl$. We fit the numerical
results as
\begin{equation}\label{e:phifit}
  \phit\phiion = c_1 + c_2 \log_{10}\,(S_0 + c_3)\,,
\end{equation}
where $c_1 = -2.89$, $c_2=2.11$, $c_3=25.3$, and
$S_0=\Sigmacl/(M_{\odot}\,{\rm pc}^{-2})$, which is plotted as the red
dashed line in \autoref{f:phiion}(c).

\subsection{Momentum Injection}\label{s:eject}

In our simulations, both neutral and ionized gas that does not
collapse to make stars is eventually ejected, carrying both mass and
momentum out of the cloud. In this section, we characterize the
efficiency of the radial momentum injection by computing net momentum
yields. We also separately assess the mean thermal and radiation
pressure forces in the radial direction, where
\begin{equation}
  \mathbf{f}_{\rm thm} = - \nabla P,  \quad
  \mathbf{f}_{\rm rad} = \frac{n_{{\rm H^0}}\sigma_{\rm H^0}}{c}
  \mathbf{F}_{\rm i} + \frac{n_{\rm H} \sigma_{\rm
      d}}{c}(\mathbf{F}_{\rm i} + \mathbf{F}_{\rm n})
\end{equation}
are the thermal and radiation pressure forces per unit volume, for
$\mathbf{F}_{\rm i}$ and $\mathbf{F}_{\rm n}$ the ionizing and
non-ionizing radiation fluxes, respectively.

\subsubsection{Net Momentum Yield}

When clouds are disrupted by feedback, the resulting outflow is
roughly spherical, so it is useful to measure the total radial
momentum of the ejected material induced by feedback. In
Section~\ref{s:dispersal}, we shall use the momentum ejected per unit
mass of stars formed, $\pyield= p_{\rm ej,final}/M_{*,\rm final}$,
which here we term the net ``momentum yield.''

\begin{figure}
  \epsscale{1.2}\plotone{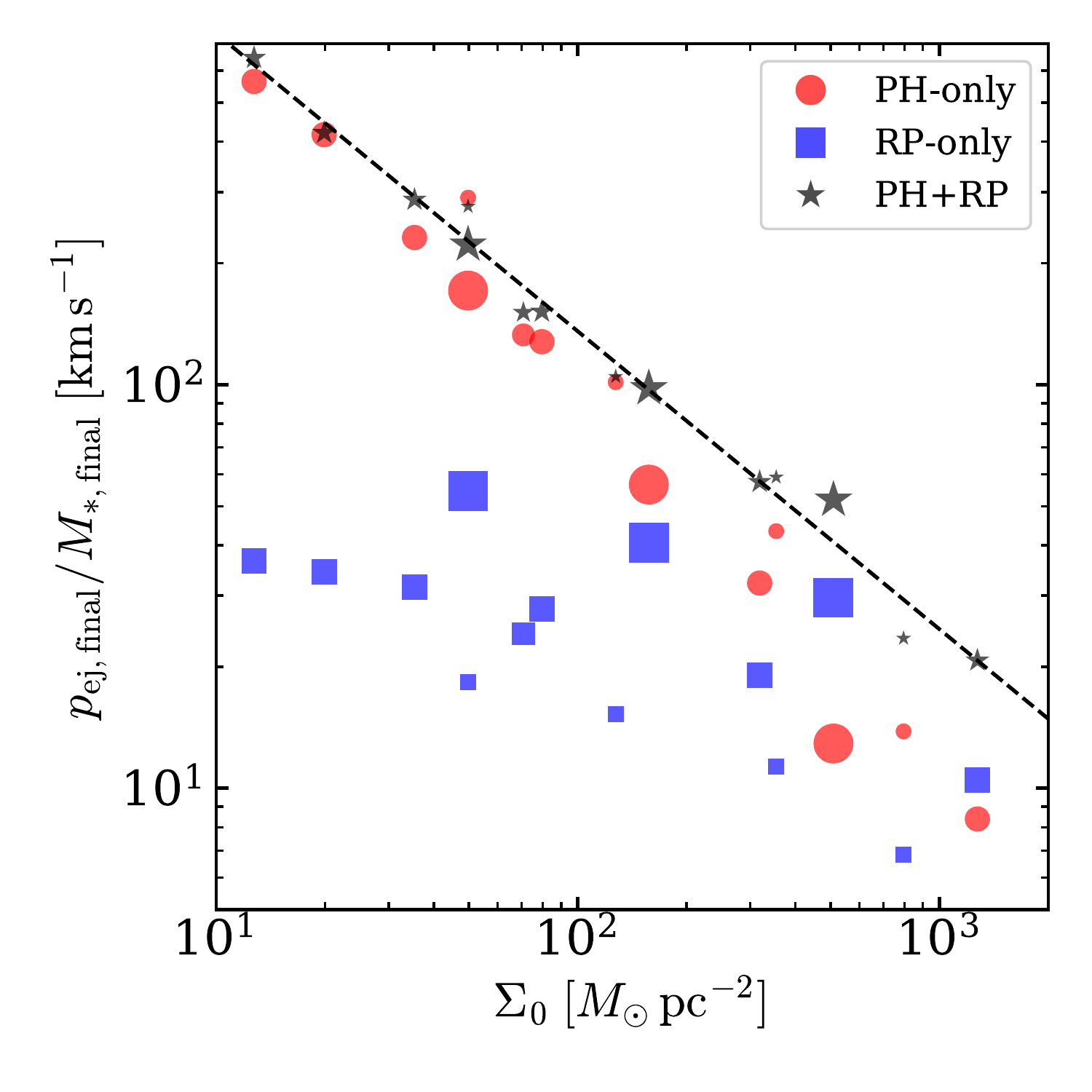}
  \caption{Total radial momentum yields
    $p_{\rm ej,final}/M_{*,\rm final}$ in the PH-only (circles),
    RP-only (squares), and PH+RP runs (stars), as a function of the
    cloud surface density $\Sigmacl$. The dashed line shows the
    power-law best-fit to the numerical results of the PH+RP runs
    (Equation~\ref{e:pyield}).}\label{f:mom}
\end{figure}

\autoref{f:mom} plots the $\pyield$ in the PH-only (circles),
RP-only (squares), and PH+RP (stars) models as a function of
$\Sigmacl$.\footnote{A small fraction of the outflow momentum is
  associated with the initial radial component of the turbulent
  outflow. For the fiducial cloud with no radiation feedback, the
  final outflow momentum is 3\% of that of the PH+RP run.} We measure
this by integrating the radial component of the momentum flux over the
outer boundary of the simulation and over all time,
$\int dt \int_{\partial V} dA \rho \mathbf{v} \cdot \hat{\mathbf{n}}
\mathbf{v} \cdot \hat{\mathbf{r}}$, where $\hat{\mathbf{n}}$ is the
normal to area $dA$. The PH-only runs have momentum yields higher than
the RP-only results except for models {\tt M1E5R05} and {\tt M1E6R25},
and close to the PH+RP results. Similarly to \autoref{f:SFE-all},
this implies that thermal pressure induced by photoionization controls
the dynamics of \HII\ regions in low surface density clouds, while
radiation pressure is more important in dense, massive clouds. For the
PH+RP and PH-only models, most of the outflow momentum is deposited in
the ionized rather than neutral gas. We find that the total momentum
yields in the PH+RP runs are well described by a power-law
relationship
\begin{equation}\label{e:pyield}
  \pyield = \frac{p_{\rm ej,final}}{M_{*,\rm final}}= 135 \kms
  \left(\frac{\Sigmacl}{10^2\Sunit}\right)^{-0.74},
\end{equation}
shown as the dashed line in \autoref{f:mom}.

We note that by adopting a light-to-mass ratio that is
time-independent, our numerical results are likely to overestimate the
true feedback strength, especially for clouds with the destruction
timescale longer than $t_{\rm SN}=3\Myr$ or $t_{\rm UV}=8\Myr$. Even
with this overestimation, our results show that the momentum injection
to the large-scale ISM from both ionizing and non-ionizing radiation
feedback is substantially smaller than that from supernovae feedback,
$p_*/m_* \sim 10^3 \kms$ \citep[e.g.,][]{kim15,kim17b}. Thus, while
momentum injection from radiation feedback is quite important on the
spatial scale of individual clouds and early life of clusters, it is
subdominant in the large-scale ISM compared to other forms of
massive-star momentum injection, and is therefore not expected to be
important in regulating either large-scale ISM pressure and star
formation rates \citep[e.g.][]{ost11,kim13} or galaxy formation
\citep[e.g.][]{naa17}.

\subsubsection{Efficiency of Radial Momentum Injection}\label{s:force}

Dynamical expansion of a spherical, embedded \HII\ region in a uniform
medium is well described by the thin-shell approximation. Force
balance $\mathbf{f}_{\rm thm} + \mathbf{f}_{\rm rad} = 0$ holds in the
interior \citep{dra11} and most of the swept-up gas is compressed into
a thin shell that expands due to the contact forces on it. The
effective momentum injection rate from gas pressure and direct
radiation pressure are $\rho_{\rm i}\ci^2 \Ai$ and $L/c$, respectively
\citep{kru09,kim16}. The former expression multiplied by a factor two
has also been applied for blister \HII\ regions, assuming that in
addition to thermal pressure the cloud back-reacts to photoevaporation
from a D-critical IF with $\rho_{\rm i}\ci^2\Ai = \dotMion\ci$
\citep[e.g.,][]{mat02,kru06}. In analytic solutions, the ionized-gas
density $\rho_i$ estimate assumes uniform density and ionization
equilibrium. For either embedded or blister \HII\ regions, this leads
to an expression for momentum injection that is given by
Equation~\eqref{e:dotNe-approx} for $\dotNe$ multiplied by $\muH \ci $
times an order-unity factor. For an idealized, dust-free, spherical,
embedded \HII\ region, the theoretical point of comparison for the gas
pressure force is
$\muH c_i^2 \left(12 \pi Q_{\rm i} \Rcl/\alphaB \right)^{1/2}$.

In reality, \HII\ regions possess many holes and ionizing sources are
widely distributed in space. The momentum injection is then expected
to be less efficient than in the idealized spherical case for the
following reasons. First, the non-spherical distribution of radiation
sources leads to momentum injection cancellation. Second, for applying
back-reaction forces, normal vectors to irregular-shaped cloud
surfaces have non-radial components with respect to the cluster
center. Third, radiation escapes through holes \citep[see
also][]{dal17,ras17}. Nevertheless, the results in Section
\ref{s:phifits} show that the numerically-computed mass-loss rates
from photoevaporation are similar to the predictions from
dimensional-analysis scaling arguments, so we may analogously expect
the radial momentum injection rate from thermal pressure to be
comparable to $\dotMion\ci$. This will be lower than the idealized
spherical momentum injection because dust absorption and escape of
radiation reduce $\Qieff$ below $\Qi$. Similarly, the radiation
pressure applied to the cloud would be reduced to $(1-f_{\rm esc})L/c$
by the escape of radiation, where
$f_{\rm esc} = (L_{\rm i,esc} + L_{\rm n,esc})/L$ is the overall
escape fraction of UV radiation; the total radiation force may be
further reduced by lack of spherical symmetry.

To determine the degree of reduction in the radial momentum injection,
we calculate the time-averaged, normalized thermal pressure force
\begin{equation}\label{e:phithm}
  \phi_{\rm thm}
  \equiv \dfrac{\int_{t_{\rm i}}^{t_{\rm f}} \int \mathbf{f}_{\rm thm}
    \cdot \hat{\mathbf{r}}\,dVdt}
{\int_{t_{\rm i}}^{t_{\rm f}} \muH \ci^2 \left(12 \pi \Qi \Rcl/\alphaB
  \right)^{1/2}\,dt}\,,
\end{equation}
\begin{equation}\label{e:phithmprime}
  \phi_{\rm thm}^{\prime}
  \equiv \dfrac{\int_{t_{\rm i}}^{t_{\rm f}}\int \mathbf{f}_{\rm
    thm} \cdot \hat{\mathbf{r}}\,dV dt}{\int_{t_{\rm i}}^{t_{\rm f}}
  \dotMion\ci \,dt}
\end{equation}
and normalized radiation pressure force
\begin{equation}\label{e:phirad}
  \phi_{\rm rad}
  \equiv \dfrac{\int_{t_{\rm i}}^{t_{\rm f}}\int \mathbf{f}_{\rm rad}
  \cdot \hat{\mathbf{r}}\,dV dt}{\int_{t_{\rm i}}^{t_{\rm f}}
  L/c\,dt}\,,
\end{equation}
\begin{equation}\label{e:phiradprime}
\phi_{\rm rad}^{\prime}
  \equiv \dfrac{\int_{t_{\rm i}}^{t_{\rm f}}\int \mathbf{f}_{\rm rad}
  \cdot \hat{\mathbf{r}}\,dV dt}{\int_{t_{\rm i}}^{t_{\rm f}}
  (1-f_{\rm esc})L/c\,dt}\,,
\end{equation}
where the range of time integration is taken as
$(t_{\rm i},t_{\rm f}) = (t_{*,0},t_{\rm ej,95\%})$.\footnote{The
  cumulative momentum injection efficiencies $\phi_{\rm thm/rad}$
  become smaller for larger $t_{\rm f}$ since most photons escape at
  late time. For example, $t_{\rm f} = t_{*,0} + t_{\rm dest}$
  increases $\phi_{\rm thm/rad}$ by $\sim 30$--$50\%$ compared to the
  case with $t_{\rm f}=t_{\rm ej,95\%} (> t_{\rm dest})$.}

In the above, we consider both the radial momentum injection from
thermal pressure and radiation pressure relative to the maximum for a
spherical shell, and relative to the actual material and radiation
momentum available.

\begin{figure}[!t]
  \epsscale{1.2}\plotone{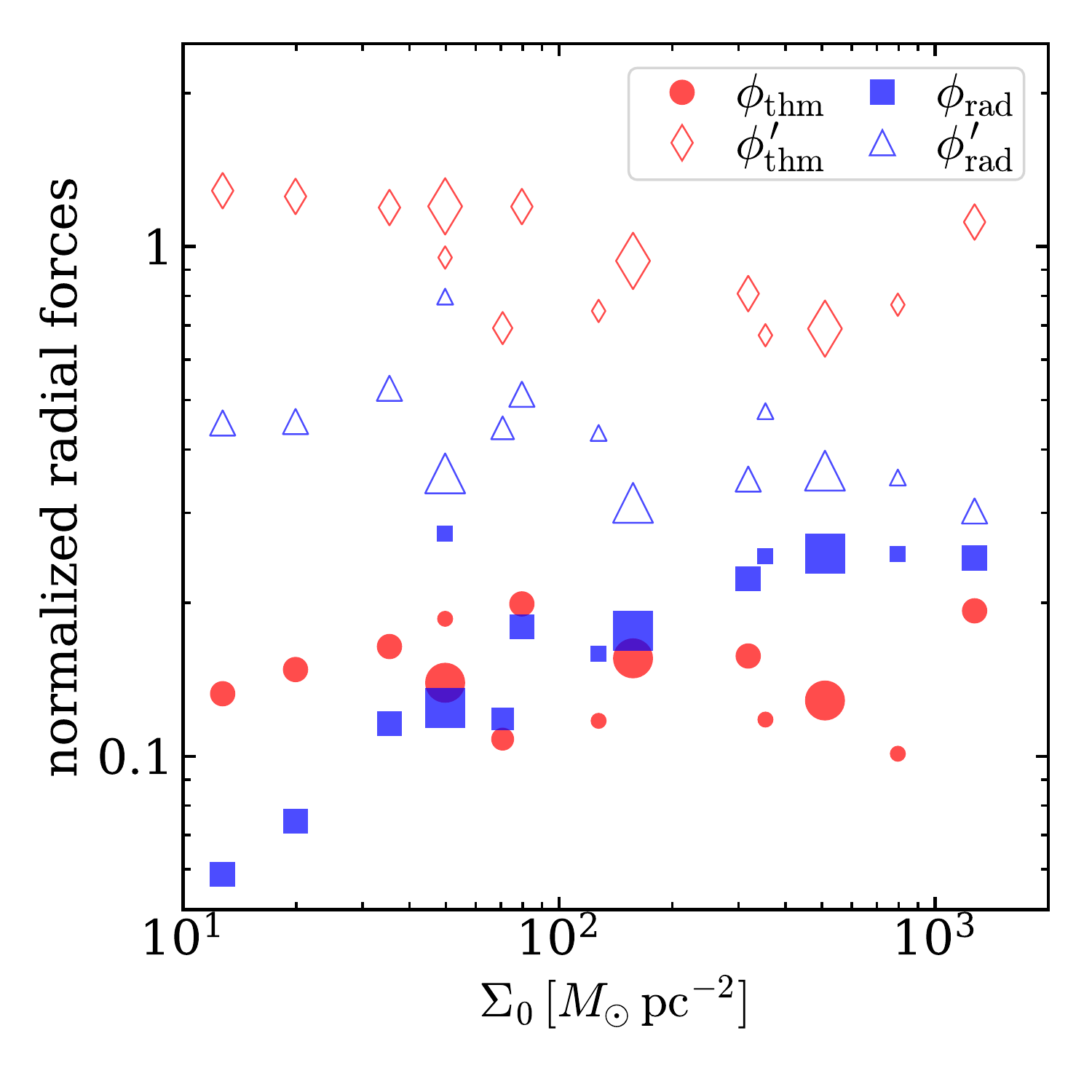}
  \caption{Average rates of the radial momentum injection due to gas
    pressure normalized by
    $\muH\ci^2\left(12 \pi \Qi \Rcl/\alphaB \right)^{1/2}$ (circles)
    and by $\dot M_{\rm ion}\ci$ (diamonds), due to radiation pressure
    normalized by $L/c$ (squares) and $(1-f_{\rm esc})L/c$ (triangles)
    for all the PH+RP models (see
    Equations~\eqref{e:phithm}--\eqref{e:phiradprime}).}\label{f:phiej}
\end{figure}

\autoref{f:phiej} plots the normalized forces $\phi_{\rm thm}$
(circles), $\phi_{\rm thm}^{\prime}$ (diamonds), $\phi_{\rm rad}$
(squares), and $\phi_{\rm rad}^{\prime}$ (triangles) averaged over the
time interval $(t_{*,0},\,t_{{\rm ej},95\%})$ as functions of
$\Sigmacl$. There is significant reduction in both the gas and
radiation pressure forces compared to the prediction for the
``spherical maximum''; the normalized gas pressure force
$\phi_{\rm thm}$ is in the range $0.10$--$0.19$ with a mean value
0.13, roughly independent of $\Sigmacl$; $\phi_{\rm rad}$ is as small
as 0.06 for model {\tt M1E5R50} and increases to 0.27 for high
surface-density clouds. The values of $\phi_{\rm thm}^{\prime}$ range
from $0.67$ to $1.26$, suggesting that the time-averaged radial
momentum injection from thermal pressure force is within $\sim 30\%$
of the naive estimate $M_{\rm ion} c_i$. The values of
$\phi_{\rm rad}^{\prime}$ are slightly larger than $\phi_{\rm rad}$
and in the range $0.3$--$0.5$, indicating that the momentum injection
by radiation pressure force is reduced by a factor of $\sim 2$--$3$ by
flux cancellation alone.\footnote{Relatively high values of
  $\phi_{\rm rad}$ and $\phi_{\rm rad}^{\prime}$ for {\tt M1E4R08}
  model are likely caused by the fact that feedback is dominated by a
  single cluster particle that accounts for 75\% of the final stellar
  mass.}

\begin{figure}[!t]
  \epsscale{1.2}\plotone{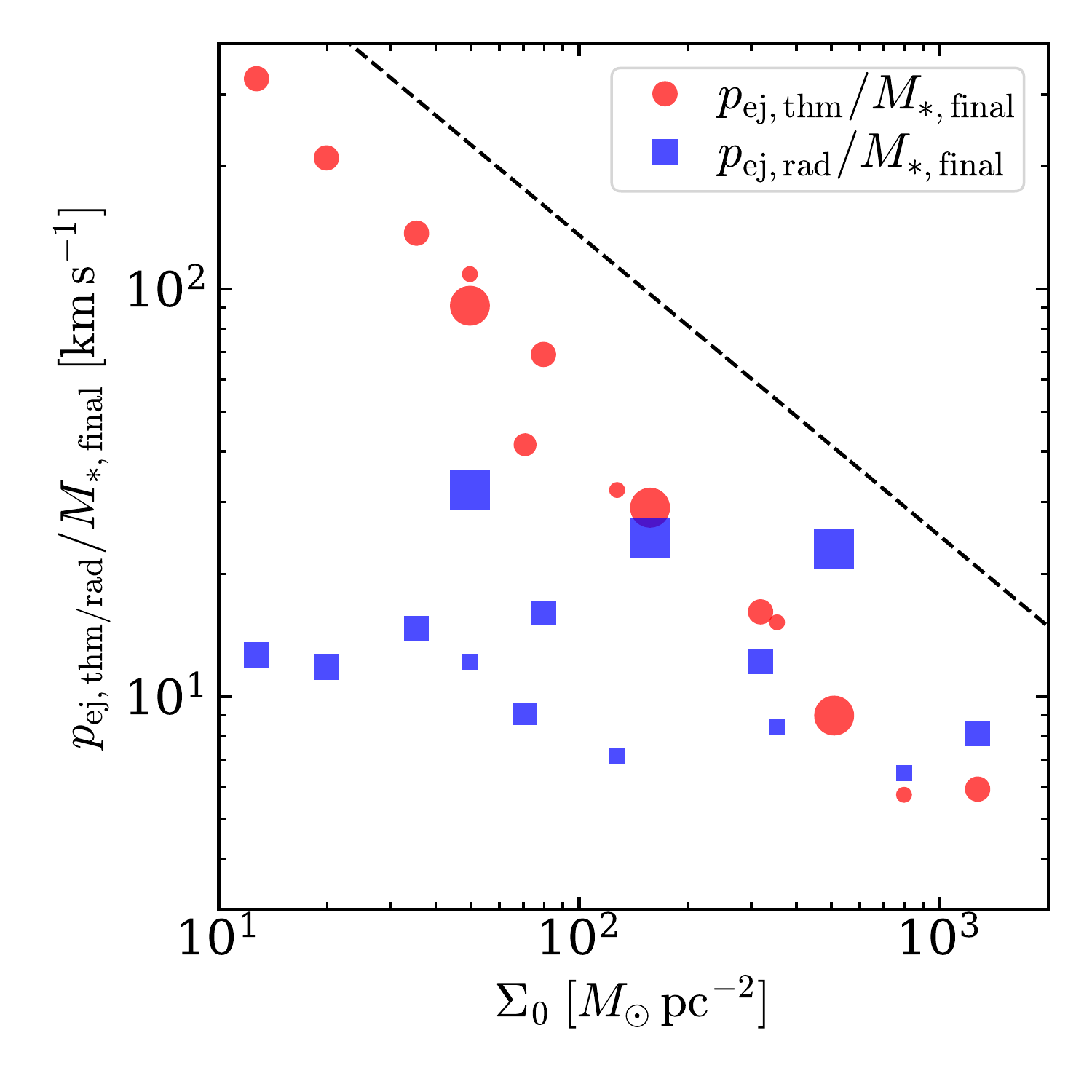}
  \caption{Injected radial momentum due to gas pressure (circles) and
    radiation pressure (squares) forces normalized by the final
    stellar mass $M_{*,{\rm final}}$ for all the PH+RP models (see
    Equations~\eqref{e:p_thmrad1} and \eqref{e:p_thmrad2}. The fit to
    the total momentum yield, Equation~\eqref{e:pyield}, is shown for
    comparison as a dashed line.}\label{f:ej-comp}
\end{figure}

Finally, we evaluate the relative importance of gas and radiation
pressure forces in radial momentum injection for the PH+RP models.
\autoref{f:ej-comp} plots the total injected momentum from gas
(circles) and radiation (squares) pressure forces per unit mass of
stars formed:
\begin{equation}\label{e:p_thmrad1}
  \dfrac{p_{\rm ej,thm}}{M_{*,{\rm final}}}
  \equiv \dfrac{\iint \mathbf{f}_{\rm thm} \cdot \hat{\mathbf{r}}\,dV
    dt} {M_{*,{\rm final}}}\,,
\end{equation}
\begin{equation}\label{e:p_thmrad2}
  \dfrac{p_{\rm ej,rad}}{M_{*,{\rm final}}}
  \equiv \dfrac{\iint \mathbf{f}_{\rm rad}
    \cdot \hat{\mathbf{r}}\,dV dt}{{M_{*,{\rm final}}}}\,.
\end{equation}
The results show trends similar to those exhibited by the total radial
momentum yield in \autoref{f:mom} for models PH vs. RP; while the
momentum injection due to gas pressure dominates in low surface
density clouds, radiation pressure takes over for clouds with
$\Sigmacl \gtrsim 500 \Sunit$. Our numerical results are consistent
with the qualitative trend in the previous analytic models
\citep{kru09,fal10,mur10,kim16} in which radiation pressure becomes
relatively more important compared to ionized-gas pressure for more
massive, high surface density clouds.

Comparison of the dashed line (Equation~\eqref{e:pyield}) and symbols
in \autoref{f:ej-comp} shows that the sum of injected momentum from
gas and radiation pressure forces $p_{\rm ej,thm} + p_{\rm ej,rad}$ is
not necessarily equal to the total momentum of outflowing gas
$p_{\rm ej,final}$. This difference arises because of the additional
centrifugal and gravitational acceleration that contributes
significantly to the latter.

\section{MODELS FOR CLOUD DISPERSAL AND SFE}\label{s:dispersal}

In this section, we use the photoevaporation rate and the momentum
yield presented in the preceding section to develop semi-analytic
models for cloud dispersal. We consider two models. First, we use the
expected physical scaling relationship between photoevaporation and
star formation efficiency (see Equation~\ref{e:Mion}) to obtain a
prediction for the SFE in the case that cloud destruction is mostly
via ionization. In the second model, we use our numerical results for
momentum injection per unit mass of stars formed to determine the SFE
necessary to eject the remaining material (both ions and neutrals)
from the cloud.

\subsection{Destruction by Photoevaporation}\label{s:dest_evap}

We first consider a situation where the gas left over from star
formation is all evaporated by photoionization. In this case, mass
conservation requires
\begin{equation}\label{e:SFE-evap0}
  1=\SFE + \epsion \,,
\end{equation}
where $\epsion =M_{\rm ion}/\Mcl=\muH\Ne/\Mcl$.

Using Equation~\eqref{e:Mion}, we express $\epsion$ as
\begin{align}\label{e:epsion}
  \epsion = \phit\phiion\left(\dfrac{\Sigmaion}{\Sigmacl}\right)\SFE^{1/2}\,,
\end{align}
where
\begin{align}\label{e:Sigmaion}
  \Sigmaion & = \muH\ci
              \left(\dfrac{\Xi}{8G\alphaB}\right)^{1/2} \\
            & = 140
              \left(\dfrac{\Xi}{5.05\times 10^{46}\,{\rm s}^{-1}\Msun^{-1}}\right)^{1/2}
              \left(\dfrac{T_{\rm ion}}{8000\,{\rm K}} \right)^{0.85}
              \Sunit\,,
\end{align}
with $\Xi = Q_{\rm i,max}/M_{*,{\rm final}}$ being the conversion
factor between stellar mass and ionizing photon output. Since
$\phiion$ is roughly constant in our numerical results,
Equation~\eqref{e:epsion} implies that the evaporation efficiency
depends linearly on the duration (relative to $t_{\rm ff}$) of
photoevaporation, $\phit$ (Equation~\eqref{e:phit}), and the square
root of the SFE, and inversely on the cloud surface density
$\Sigmacl$. The characteristic surface density $\Sigmaion$
  can be regarded as a constant for most of the models since
  $M_{*,{\rm final}} > 10^3 \Msun$, while it varies sensitively with
  $M_{*,{\rm final}}$ in models with $M_{*,{\rm final}} < 10^3 \Msun$,
  as described in Section \ref{s:RHScheme}. The dependence of
$\epsion$ on $\SFE^{1/2}$ (rather than $\SFE$) is caused by shielding
within the IBSV, which makes photoevaporation inefficient. Most
photons are used up in offsetting recombination within the \HII\
region, rather than in eroding neutral gas to create new ions at IFs.

We solve Equations~\eqref{e:SFE-evap0} and \eqref{e:epsion} for $\SFE$
to obtain
\begin{equation}\label{e:SFE-evap2}
  \SFE = \left(\frac{2\xi}{1 + \sqrt{1 + 4 \xi^2}}\right)^2\,,
\end{equation}
where $\xi\equiv \Sigmacl/(\phit\phiion\Sigmaion)$. In the limit
$\xi\ll 1$,
$\SFE \approx \xi^2 = [\Sigmacl/(\phit\phiion\Sigmaion)]^2$, while in
the limit $\xi \gg 1$,
$\SFE \approx 1- 1/\xi= 1- \phit\phiion\Sigmaion/\Sigmacl$. The
limiting behavior shows that photoevaporation becomes very efficient
for destroying clouds when $\Sigmacl \lesssim \Sigma_{\rm ion}$.

Inserting the fit for $\phit\phiion$ of Equation~\eqref{e:phifit} in
Equation~\eqref{e:SFE-evap2} allows us to calculate the net SFE as a
function of $\Sigmacl$. The resulting $\SFE$ and $\epsion$ are
compared to the numerical results as dashed lines in
\autoref{f:eps-all}(a) and (c). For low-surface density and massive
clouds with $\epsion \gtrsim 0.7$, the semi-analytic $\SFE$ agrees
with the numerical results extremely well. The agreement is less good,
however, for high-surface density
($\Sigmacl\gtrsim 300 \Msun\;\rm pc^{-2}$) or low-mass
($\Mcl=10^4 \Msun$) clouds for which neutral gas ejection cannot be
ignored; we address this next.

\subsection{Destruction by Dynamical Mass Ejection}\label{s:dest_ej}

We consider the general case in which radiative feedback from star
formation exerts combined thermal and radiation pressure forces on the
surrounding gas. The gas is ejected from the cloud, quenching further
star formation. In our simulations, the momentum ejected from the
computational domain also includes a small contribution from the
initial turbulence, which ejects a mass fraction
$\epstu=M_{\rm ej,turb}/\Mcl \sim 0.1$ when the cloud is initially
marginally bound.

Let $v_{\rm ej}$ denote the characteristic ejection velocity of the
outgoing gas, and let $\pyield$ denote the momentum per stellar mass
injected by feedback. The total momentum of the ejected gas can be
written as
\begin{equation}\label{e:DME}
  p_{\rm ej,tot} = (1 - \SFE)\Mcl v_{\rm ej} =(\pyield)\SFE \Mcl +
  \epstu \Mcl v_{\rm ej}\,.
\end{equation}
We divide the both sides of Equation~\eqref{e:DME} by
$\Mcl v_{\rm ej}$ to obtain
\begin{equation}\label{e:SFE-ej}
  \SFE = \dfrac{1 - \epstu}{1 + (\pyield)/v_{\rm ej}}\,.
\end{equation}

From the results of model {\tt M1E5R20\_nofb} (see
\autoref{t:result}), we take $\epstu=0.13$.\footnote{Similar to our
  {\tt M1E5R20\_nofb} model, \citet{ras16} ran a suite of
  `no-feedback' simulations and found that clouds with
  $\alphaviro = 2$ have roughly the same turbulence ejection
  efficiencies $0.10 < \epstu < 0.15$ regardless of $\Mcl$ and
  $\Sigmacl$.} For $v_{\rm ej}$, we take the mass-weighted outflow
velocity
$(\epsejneu v_{\rm ej,neu} + \epsejion v_{\rm ej,ion})/\epsej$, which
is found to be roughly constant for fixed $\Mcl$ with mean values
$15$, $23$, $29 \kms$ for $\Mcl = 10^4$, $10^5$, $10^6 \Msun$,
respectively. Equation~\eqref{e:SFE-ej} gives $\SFE$ as functions of
$\Mcl$ and $\Sigmacl$ once $\pyield$, $v_{\rm ej}$ and $\epstu$ are
specified. A fit to the momentum yield $\pyield$ from our numerical
results as a function of $\Sigmacl$ is given in Equation
\eqref{e:pyield}. Altogether, Equation~\eqref{e:SFE-ej} then gives
$\SFE$ as a function of $\Mcl$ and $\Sigmacl$. For
$\pyield \gg v_{\rm ej}$ and $\epstu \ll 1$, the predicted scaling
dependence is
$\SFE \sim v_{\rm ej}/(\pyield) \propto v_{\rm ej} \Sigmacl^{3/4}$.

\autoref{f:eps-all} overplots as solid lines the resulting (a) net
SFE $\SFE$, (b) ejection efficiency $\epsej = 1- \SFE$, (c)
photoevaporation efficiency $\epsion$ calculated from
Equation~\eqref{e:epsion}, and (d) neutral ejection efficiency
$\epsejneu = \epsej - \epsion$. The line thickness indicates the
initial cloud mass. Overall, our semi-analytic model for the dynamical
mass ejection reproduces the net SFE of the simulations fairly well,
with the average difference of $0.03$. The model also explains the
increasing tendency of $\SFE$ with increasing $\Mcl$. This weak
dependence of the semi-analytic $\SFE$ on $\Mcl$ comes directly from
$v_{\rm ej}$, indicating that stronger feedback is required to unbind
gas in a more massive cloud \citep[e.g.,][]{kim16,rah17}. A drop in
the photoevaporation efficiency with decreasing $\Sigmacl$ for
$\Mcl = 10^4 \Msun$ and $\Sigmacl \lesssim 10^2 \Sunit$ is caused by
the steep dependence of $\Xi$ on $M_{*,{\rm final}}$ below
$10^3\Msun$, as mentioned above.

\section{SUMMARY AND DISCUSSION}\label{s:sum}

In this work, we have used our new implementation of adaptive ray
tracing in the \Athena\ magnetohydrodynamics code to simulate
star cluster formation and the effects of radiation feedback on
turbulent GMCs. We consider a suite of clouds (initially marginally
bound) with mass $\Mcl=10^4$--$10^6 \Msun$ and radius
$\Rcl=2$--$80 \pc$; the corresponding range of the surface density is
$\Sigmacl \approx 13$--$1300 \Sunit$. The primary goals of the current
paper are to understand the role of (ionizing and non-ionizing) UV
radiation feedback in controlling the net SFE and GMC lifetime and to
assess the relative importance of photoionization and radiation
pressure in various environments. We augment and compare our numerical
results with semi-analytic models.

Our main findings are summarized as follows.

\subsection{Summary}

\begin{enumerate}

\item {\textit{Evolutionary stages}}

  All clouds in our simulations go through the following
  evolutionary stages (Figure \ref{f:proj-slice}), with some overlap
  and varying duration. (1) The initial turbulence creates filaments,
  and within these denser clumps condense. Some clumps undergo
  gravitational collapse and form star particles representing
  subclusters. (2) Individual \HII\ regions form around each
  subcluster, growing towards directions of low optical depth until
  they merge and break out of the cloud. Simultaneously, both low- and
  high-density gas is accelerated away from the radiation sources due
  to thermal and radiation pressures. (3) At late stages of evolution
  the brightest EM features are small globules, pillars, and ridges
  marking individual IFs at the periphery of the \HII\ region, quite
  similar to observed clouds (\autoref{f:nesq}). (4) At the end of
  the simulation, all of the gas has either collapsed into stars or
  been dispersed, flowing out of the computational domain.

\item {\textit{Timescales}}

  Cloud destruction takes $\sim 2$--$10\Myr$ after the onset of
  massive star formation feedback (\autoref{f:timescales}(a)). In
  units of the freefall time, the destruction timescale is in the
  range $0.6 < t_{\rm dest}/\tffo < 4.1$ and systematically increases
  with $\Sigmacl$ (\autoref{f:timescales}(b)). The timescale for
  star formation is comparable to or somewhat smaller than the
  destruction timescale. The effective gas depletion timescale ranges
  from $250\Myr$ to $2.7\Myr$, sharply decreasing with increasing
  $\Sigmacl$.

\item {\textit{Star formation and mass loss efficiencies}}

  We examine the dependence on the cloud mass and surface density of
  the net SFE $\SFE$, photoevaporation efficiency $\epsion$, and mass
  ejection efficiency $\epsej$ (\autoref{f:eps-all}). The SFE
  ranges over $\SFE= 0.04$--$0.51$, increasing strongly with the
  initial surface density $\Sigmacl$ while increasing weakly with the
  initial cloud mass $\Mcl$. Photoevaporation accounts for more than
  70\% of mass loss for clouds with $\Mcl \gtrsim 10^5\Msun$ and
  $\Sigmacl \lesssim 10^2 \Sunit$. The ejection of neutral gas mass by
  thermal and radiation pressures also contributes in quenching
  further star formation in low-mass and high-surface density clouds.
  The comparison of the net SFE among the models in which we turn on
  and off photoionization and radiation pressure suggests that
  photoionization is of greater importance in destroying GMCs in
  normal disk galaxies, whereas radiation pressure is more effective
  in regulating star formation in dense, massive clouds
  (\autoref{f:SFE-all}; see also \autoref{f:ej-comp}). This
  controlled experiment also demonstrates that the combined effects of
  photoionization and radiation pressure do not work in a simply
  additive manner in suppressing star formation.

\item {\textit{Photoevaporation}}

  The photoevaporation rate $\dotNe$ (the number of free electrons
  produced per unit time) at IFs within the cloud is much smaller than
  the total number of ionizing photons absorbed by hydrogen per unit
  time $\Qieff$ throughout the cloud (\autoref{f:hst-dot-Mevap}).
  Most of the ionizations instead offset recombinations in diffuse gas
  throughout the cloud. The time-averaged hydrogen absorption fraction
  $\langle f_{\rm ion} \rangle = \langle \Qieff/\Qi \rangle$ ranges
  from $0.22$ to $0.50$, while the time-averaged shielding factor
  $\langle q \rangle = \langle \Qieff/\dotNe \rangle$ ranges from $61$
  to $2950$ (\autoref{t:result}). Although dense, compact clouds
  tend to have a high hydrogen absorption fraction, they have a high
  shielding factor and hence inefficient photoevaporative mass loss.
  Assuming that the area of the IFs and thickness of the shielding
  layer scale with the dimensions of the initial cloud, we derive and
  calibrate expressions for the photoevaporation rate
  (Equations~\eqref{e:dotNe-approx}, \eqref{e:phiion};
  \autoref{f:phiion}a).

\item {\textit{Outflow acceleration and properties}}

  We perform a detailed analysis of the momentum injection processes
  that are responsible for cloud disruption and mass loss. The total
  radial momentum yield (momentum per stellar mass formed) of
  outflowing gas ranges over $\sim 20$--$400 \kms$
  (\autoref{f:mom}), scaling as $\pyield \propto \Sigmacl^{-0.74}$
  (Equation~\eqref{e:pyield}). The time-averaged total radial gas
  pressure force is smaller than the dustless, spherical case by a
  factor of $\sim 5$--$10$ due to momentum cancellation and escape of
  radiation, but is within $30\%$ of $\dotMion \ci$, where $\dotMion$
  and $\ci$ refer to the mass evaporation rate and the sound speed of
  the ionized gas, respectively (\autoref{f:phiej}). This is
  consistent with expectations for both internal pressure forces
  within the ionized medium and the combined thermal pressure and
  recoil forces on neutral gas at IFs, both of which scale as
  $\nion \ci^2 \Ai \sim \dotMion \ci$. Similarly, the time-averaged
  radial force from radiation pressure is much reduced below the naive
  spherical expectation to $\sim 0.08$--$0.27 L/c$ because of flux
  cancellation and photon escape. Although the overall momentum
  injection by radiation feedback is less efficient in realistic
  turbulent clouds than suggested by analytic predictions based on
  spherical \HII\ region expansion in smooth clouds
  \citep[e.g.][]{kru09,mur10,fal10,kim16}, the ratio of radiation
  pressure forces to gas pressure forces increases for massive, high
  surface density clouds (\autoref{f:ej-comp}), consistent with
  expectations from these previous studies. The mean outflow velocity
  of ionized gas is mildly supersonic with
  $v_{\rm ej,ion} \sim 18$--$36\,{\rm km\,s}^{-1}$, while that of
  neutral gas is $v_{\rm ej,neu} \sim 6$--$15\kms$, at about
  $0.8$--$1.8$ times the escape velocity of the initial cloud
  (\autoref{f:vej} and \autoref{t:result}).

\item {\textit{Semi-analytic models}}

  Based on our analyses of mass loss processes, we develop simple
  semi-analytic models for the net SFE, $\SFE$, and photoevaporation
  efficiency, $\epsion$, as limited by radiation feedback in
  cluster-forming clouds. The predicted $\epsion$ is proportional to
  $\SFE^{1/2} \Sigmacl^{-1}$ (Equation~\eqref{e:epsion}), with an
  order-unity dimensionless coefficient $\phit \phiion$ that we
  calibrate from simulations (Equation~\eqref{e:phifit},
  \autoref{f:phiion}). When photoevaporation is the primary agent
  of cloud destruction, the net SFE depends solely on the gas surface
  density (Equation~\eqref{e:SFE-evap2}). The resulting predictions
  for $\SFE$ and $\epsion$ agree well with the numerical results for
  massive ($\ge 10^5 \Msun$) clouds (\autoref{f:eps-all}(a) and
  (c)). In low-mass clouds ($10^4 \Msun$), the back-reaction to
  ionized-gas pressure is effective at IFs, and these clouds lose
  $30-50\%$ of their initial gas mass as neutrals. Allowing for both
  ionized and neutral gas in outflows and assuming that the ejection
  velocity is constant for fixed cloud mass, with total momentum
  injection calibrated from our simulations (Equation~\ref{e:pyield}),
  the predicted net SFE (Equation~\eqref{e:SFE-ej}) has a weak
  dependence on the cloud mass, consistent with our numerical results
  (\autoref{f:eps-all}).

\begin{figure*}[!ht]
  \epsscale{1.}\plotone{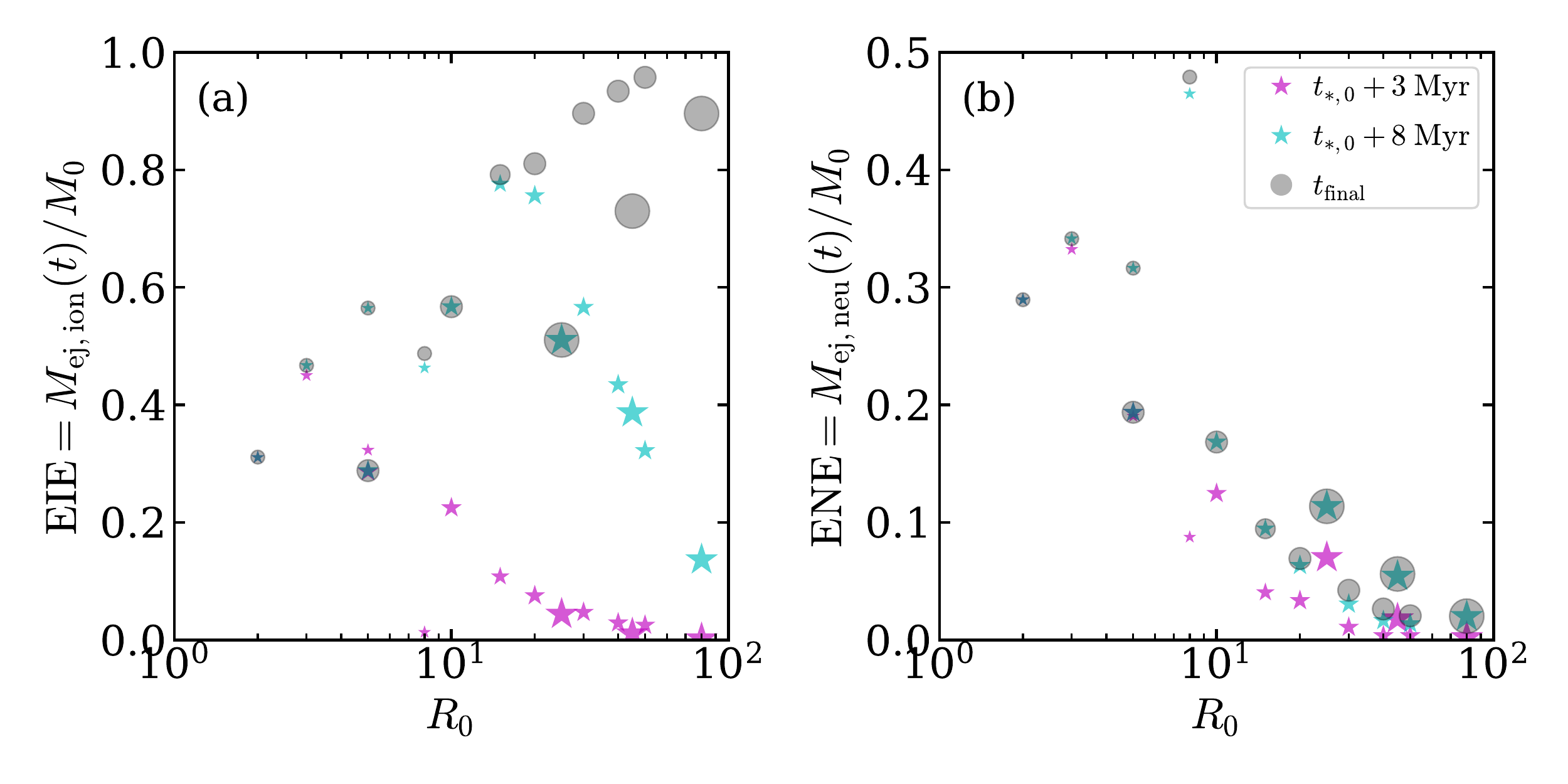}
  \caption{(a) Ejected ion efficiency (EIE) and (b) ejected neutral
    efficiency (ENE) at times $t_{*,0} + t_{\rm SN}$ (magenta),
    $t_{*,0} + t_{\rm UV}$ (cyan), and $t_{\rm final}$ (gray) as
    functions of the initial cloud radius $\Rcl$.}\label{f:tSN-tUV}
\end{figure*}

\end{enumerate}

\subsection{Discussion}\label{s:discuss}

It is interesting to compare our results with those of previous
theoretical studies on star-forming GMCs with UV radiation feedback.
Our net SFE and relative role of photoionization to radiation pressure
are in qualitative agreement with the predictions of \citet{kim16},
which adopted the idealizations of instantaneous star formation and
spherical shell expansion (see also \citealt{fal10,mur10,rah17}).
However, the minimum SFE required for cloud disruption by ionized-gas
pressure found by \citet{kim16} ($\sim 0.002$--$0.1$ for
$20 \Sunit \leq \Sigmacl \leq 10^2 \Sunit$, see their Figure~11) is
smaller than what we find here. Turbulence-induced structure can lead
to higher SFE for several reasons. First, a fraction of the photons
can easily escape through low-density channels, without either
ionizing gas or being absorbed to impart momentum. Second, low-density
but high pressure ionized gas can also vent through these channels,
reducing the transfer of momentum from photoevaporated to neutral gas.
Third, turbulence increases the mass-weighted mean density and
therefore the recombination rate of ionized gas, so that a higher
luminosity is required to photoevaporate gas.

In this paper, we also find that radiation pressure becomes
significant at $\Sigmacl \sim 200$--$800\Sunit$, lower for higher-mass
clouds. This transition value of surface density for low-mass clouds
is somewhat higher than $\Sigmacl \sim 100 \Sunit$ in simple spherical
models \citep[e.g.][]{kru09,mur10,fal10,kim16}. These spherical models
predicted that the expansion of an \HII\ region is dominated by
radiation pressure force at least in the early phase of expansion
\citep[e.g.,][]{kru09,kim16}. Even for the highest-$\Sigmacl$ models,
however, we find that the volume-integrated radiation pressure force
($\int \mathbf{f}_{\rm rad} \cdot \mathbf{\hat{r}}\,dV$) in the radial
direction tends to be smaller than that of gas pressure force
($\int \mathbf{f}_{\rm thm} \cdot \mathbf{\hat{r}}\,dV$) in the early
phase of star formation, and that this tendency is reversed only at
late time. This is due to the strong cancellation of radiation on a
global scale, and should thus not be interpreted as evidence for
radiation pressure being subdominant. Rather, radiation pressure plays
a greater role in controlling the dynamics of sub \HII\ regions
surrounding individual sources.

The outflow momentum yield of $\pyield \sim20$--$400\kms$ by radiation
feedback found from our simulations can be compared with those
produced by other types of feedback. For example, the momentum yield
of protostellar outflows is estimated as $\sim 40 \kms$
\citep[e.g.,][]{mat00}, while recent numerical work on SNe-driven
flows found a momentum yield of $\sim 1000$--$3000 \kms$, weakly
dependent on the background density and stellar clustering
\citep[e.g.,][]{kim15,kim17b}. These feedback mechanisms are expected
to be important at different scales
\citep[e.g.,][]{fal10,kru14,mat15}. The momentum injection by
protostellar outflows plays a key role in driving turbulence and
regulating star formation in small-scale clumps before massive stars
form \citep[e.g.,][]{nak14}, and SNe are capable of driving ISM
turbulence to regulate galaxy-wide star formation and possibly winds
from low-mass galaxies \citep[e.g.,][]{kim17b}. Momentum injection by
UV radiation is important to controlling dynamics at intermediate, GMC
scales.

In this paper we have focused exclusively on the effects of radiation
feedback, and as we neglect aging of stellar populations over the
whole cloud lifetime, our mass-loss efficiencies place an upper limit
to the actual damage UV radiation can cause. To estimate lower limits
to the destructive effects of radiation feedback, we can consider what
UV feedback would be able to accomplish over two shorter intervals:
$t_{\rm SN} = 3\Myr$, the minimum lifetime of most massive stars
(i.e., the time when first supernova is expected to occur after the
epoch of the first star formation); and $t_{\rm UV}=8\Myr$, the time
scale on which UV luminosity decays. \autoref{f:tSN-tUV} shows the
cumulative (a) ejected ion efficiency
${\rm EIE} = M_{\rm ej,ion}/\Mcl$ and (b) ejected neutral efficiency
${\rm ENE} = M_{\rm ej,neu}/\Mcl$ as functions of the initial cloud
radius $\Rcl$ at $t_{*,0} + t_{\rm SN}$ (magenta),
$t_{*,0} + t_{\rm UV}$ (cyan), and $t_{\rm final}$ (gray). In the
case of compact, high-surface density clouds with short freefall
times, the mass-loss efficiencies at $t_{*,0}+t_{\rm SN}$ are close to
the final values at $t_{\rm final}$, suggesting that UV radiation is
expected to clear out most of gas prior to the first supernova. For
large, diffuse clouds with long freefall times, the mass ejection
efficiencies at $t_{*,0} +t_{\rm SN}$ and $t_{*,0} +t_{\rm UV}$ are
significantly smaller than the final values.
For these diffuse clouds with long freefall timescales, a complete
assessment of star formation efficiency and cloud destruction will
have to include the effects of supernovae as well as radiation
feedback. While some of the supernova energy escapes easily through
low-density channels \citep[e.g.][]{rog13,wal15}, and supernovae may
undermine radiation feedback by compressing overdense structures,
supernova feedback likely aids in destroying star-forming clouds
overall \citep[e.g.][]{gee16}. In principle, shocked stellar winds may
also be important over the same period as radiation feedback is
active, although recent studies have raised doubts about its
effectiveness as this hot gas can easily leak through holes or mix
with cool gas \citep{har09,ros14}.

In this work, we have considered only unmagnetized, marginally bound
clouds with $\alphaviro=2$, while observed GMCs may have a range of
virial values \citep[e.g.,][]{hey09,rom10,miv17}. Preliminary
simulations we have done reveal that the net SFE decreases from $0.31$
to $0.02$ as $\alphaviro$ is varied from $0.5$ to $5$ for the fiducial
mass and size. Clouds with large initial $\alphaviro$ tend to form
stars less efficiently since turbulence unbinds a larger fraction of
gas and reduces the gas mass in collapsing structures, as has been
reported by recent simulations
\citep[e.g.,][]{dal13,ber15,how16,ras16,dal17}. For clouds that are
initially strongly-bound, an initial adjustment period leads to
smaller cloud with order-unity $\alpha$, in which the subsequent
evolution is similar to clouds with $\alphaviro \sim 1$ \citep{ras16}.
Magnetization is likely to increase the effectiveness of radiation
feedback, as it reduces the density inhomogeneity that limits
radiation pressure effects, and may also help to distribute energy
into neutral gas \citep{gen12}.

Finally, we comment on the resolution of our numerical models. Our
standard choice of grid resolution ($N_{\rm cell} = 256^3$) and domain
size ($L_{\rm box}=4\Rcl$) is a compromise between computational time
and accuracy, and this choice of moderate resolution enabled us to
extensively explore parameter space. While the present simulations can
capture the dynamics of cluster-forming gas and outflowing gas on
cloud scales with reasonable accuracy, they do not properly resolve
compressible flows in high-density regions where sink particles are
created. Our models may overestimate star formation rates as all the
gas that has been accreted onto sink particles is assumed to be
converted to stars \citep[cf.][]{how16,how17}, although a converging
trend of $\SFE$ with resolution (see Appendix \ref{s:app1}) suggests
that the final SFE may be primarily controlled by photoevaporation and
injection of momentum in moderate density gas at large scales.
Adaptive mesh refinement simulations can be used to address this and
other resolution-related questions.

\acknowledgements We thank the anonymous referee for constructive
comments on the manuscript. J.-G.K. would like to thank Michael
Grudi{\'c}, Takashi Hosokawa, Ralph Pudritz, and Benny Tsang for
stimulating discussions. J.-G.K. acknowledges financial support from
the National Research Foundation of Korea (NRF) through the grant
NRF-2014-Fostering Core Leaders of the Future Basic Science Program.
The work of W.-T.K. was supported by the National Research Foundation
of Korea (NRF) grant funded by the Korean government (MEST)
(No.~3348-20160021). The work of E.C.O. was supported by grant
AST-1713949 from the U.S. National Science Foundation. The computation
of this work was supported by the Supercomputing Center/Korea
Institute of Science and Technology Information with supercomputing
resources including technical support (KSC-2017-C3-0029), and the
PICSciE TIGRESS High Performance Computing Center at Princeton
University.

\software{Athena \citep{sto08}, yt \citep{tur11}, numpy \citep{van11},
  matplotlib \citep{hun07}, IPython \citep{per07}, pandas
  \citep{mck10}, ParaView \citep{aya15}.}

\appendix

\section{Resolution Study of the Fiducial Model}\label{s:app1}

\begin{figure}[!t]
  \epsscale{1.2}\plotone{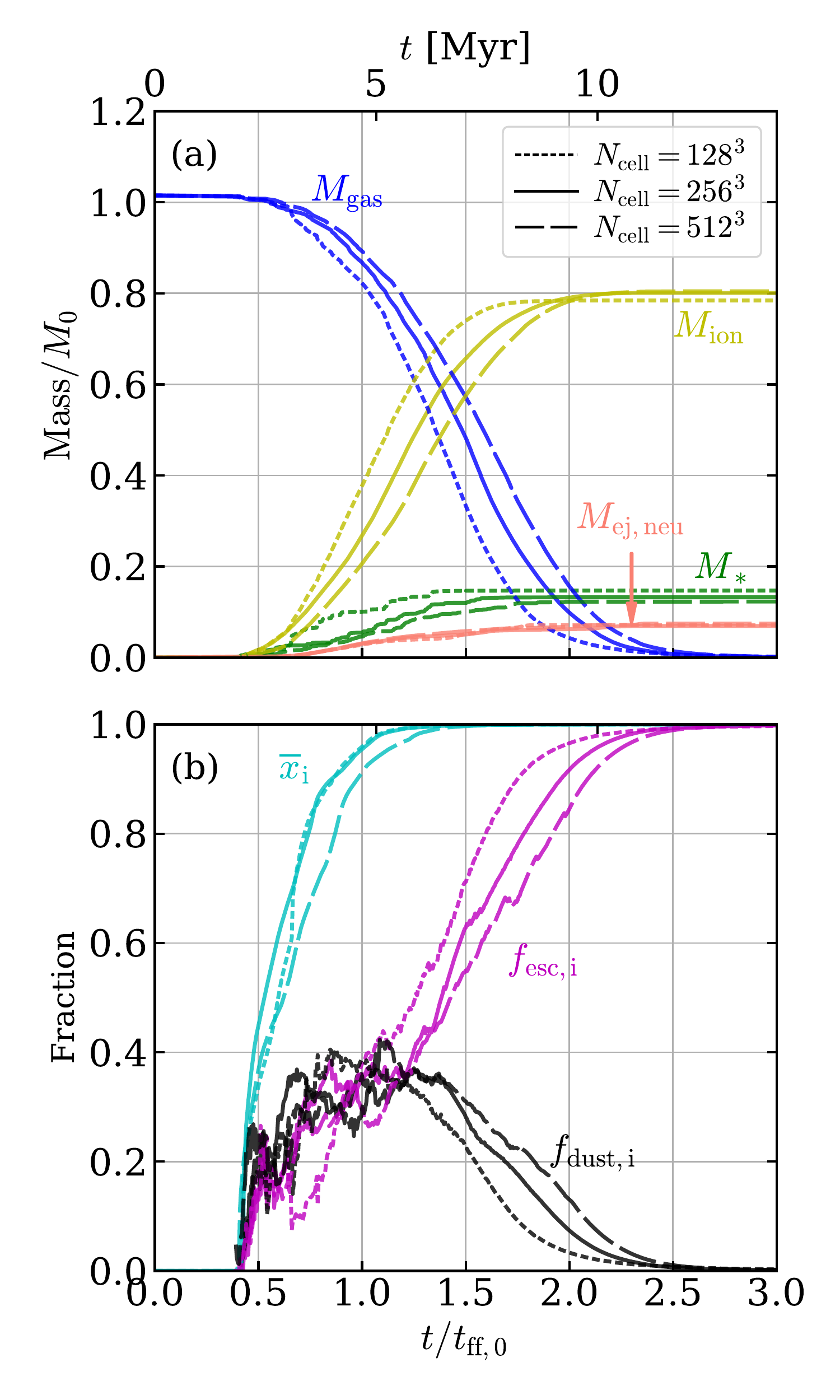}
  \caption{Evolutionary histories of key quantities for the fiducial
    cloud model ($\Mcl=10^5\Msun$ and $\Rcl = 20\pc$) at varying
    resolution: $N_{\rm cell} = 128^3$ (short dashed), $256^3$
    (solid), and $512^3$ (long dashed). (a) The total gas mass
    $M_{\rm gas}$ in the simulation volume (blue), the stellar mass
    $M_*$ (green), the ejected neutral gas mass $M_{\rm ej,neu}$
    (salmon), and the mass of the photoevaporated gas $\Mion$
    (yellow). (b) The volume fraction of the ionized gas
    $\overline{x}_{\rm i}$ (cyan), the fraction of ionizing radiation
    absorbed by dust $f_{\rm dust,i}$ (black), and the escape
    fractions of ionizing photons $f_{\rm esc,i}$
    (magenta).}\label{f:hst-res}
\end{figure}

\begin{figure*}[!t]
  \epsscale{1.05}\plotone{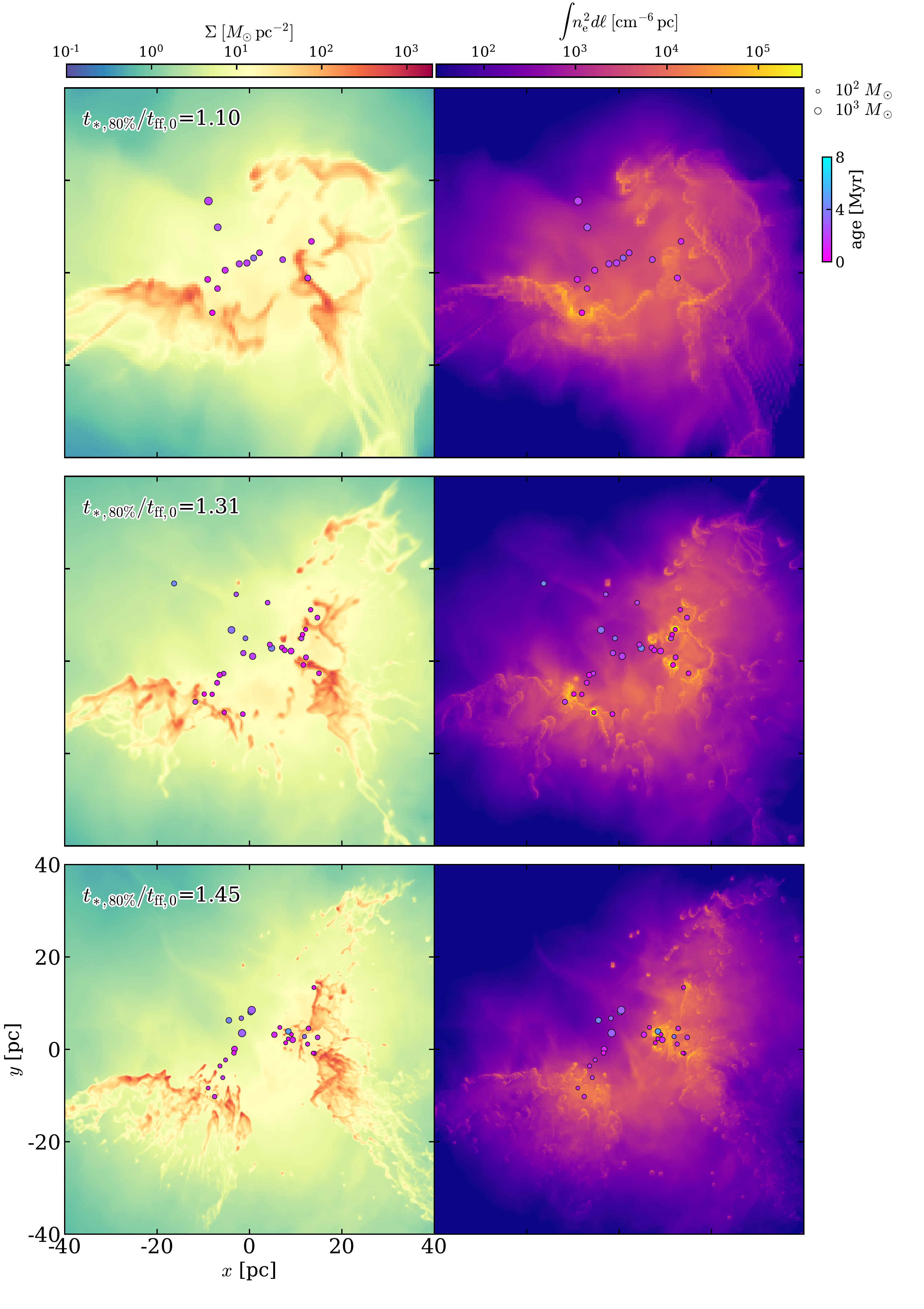}
  \caption{Snapshots of the gas surface density (left) and emission
    measure (right) in the fiducial model when 80\% of the final
    stellar mass has been assembled. The top, middle, and bottom
    panels correspond respectively to the run with
    $N_{\rm cell} = 128^3$ at $t/t_{\rm ff,0}=1.10$, $256^3$ at
    $t/t_{\rm ff,0}=1.31$, and $512^3$ at $t/t_{\rm ff,0}=1.45$. The
    projected positions of star particles are shown as circles, with
    their color corresponding to age.}\label{f:proj-512}
\end{figure*}

The high computational cost required for radiation transfer,
especially for models involving numerous point sources, prevents us
from running all our simulations at very high resolution. To study how
our numerical results depend on the grid size, we run the fiducial
model with $\Mcl=10^5 \Msun$ and $\Rcl = 20\pc$ at three different
spatial resolutions with $N_{\rm cell} = 128^3$ ({\tt M1E5R20\_N128}),
$256^3$ ({\tt M1E5R20}), and $512^3$ ({\tt M1E5R20\_N512}). Here we
compare the results of these models.

\autoref{f:hst-res} plots the temporal evolution of various volume-
or surface-integrated quantities from {\tt M1E5R20\_N128} (short
dashed), {\tt M1E5R20} (solid), {\tt M1E5R20\_N512} (long dashed). It
is clear that all quantities exhibit qualitatively similar behavior
with time, and the final values of key quantities such as stellar mass
and photoevaporated gas mass listed in \autoref{t:result} are
numerically quite close and follow a converging trend as the
resolution increases. For example, the increment in the duration of
star formation $t_{\rm SF}$ from the $128^3$ to $256^3$ runs is
$\sim 25\%$, which is reduced to $\sim 17\%$ from the $256^3$ to
$512^3$ runs. This resolution-dependent $t_{\rm SF}$ is due primarily
to the fact that the minimum sink particle mass as well as the
physical size of the control volume (or the effective area through
which gas accretes) are proportional to $\Delta x$ (or $\Delta x^2$),
so that the stellar mass grows more rapidly and in a more discrete
fashion at coarser spatial resolution. Despite this difference, the
net SFE of $0.15$, $0.13$, and $0.12$ for $N_{\rm cell} = 128^3$,
$256^3$, and $512^3$, respectively, is almost converged.

\autoref{f:proj-512} compares sample snapshots of the total
(neutral + ionized) gas surface density and the emission measure
projected along the $z$-axis when 80\% of the final stellar mass has
been formed (at $t/t_{\rm ff,0}=1.10, 1.31, 1.45$ for
$N_{\rm cell} = 128^3$, $256^3$, and $512^3$, respectively). The
higher-resolution model exhibits filaments, pillars, and bright-rimmed
globules in greater detail, but the overall morphologies of gas and
star particle distributions are very similar. Therefore, we conclude
that the results based on $N_{\rm cell}=256^3$ presented in the paper
provide quantitatively reasonable estimates to the converged results.

\section{Net Star Formation Efficiency Regulated by Radiation-Pressure
  Feedback}\label{s:app2}

\begin{figure}
  \epsscale{1.2}\plotone{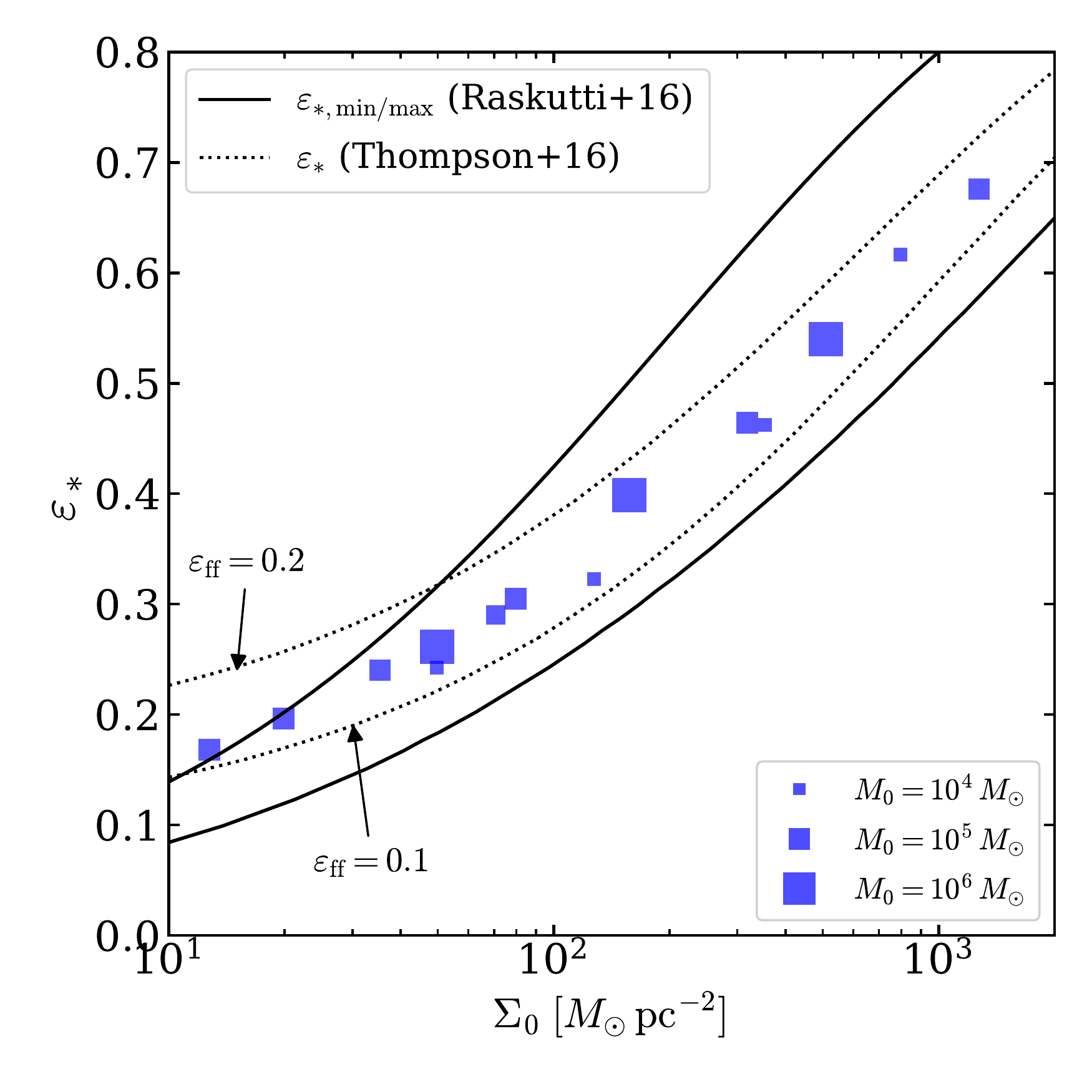}
  \caption{Comparison of the net SFE from the RP-only simulations
    (squares) with those from the theoretical predictions (lines). The
    solid lines show $\varepsilon_{*,\rm min}$ and
    $\varepsilon_{*,\rm max}$ based on the analytic model of
    \citet{ras16}. The predictions of \citet{tho16} with
    $\varepsilon_{\rm ff} = 0.1, 0.2$ are given as dotted lines. For
    both models we adopt $\sigma_{\ln \Sigma^c}=1.4$ based on the
    typical variance in our simulations.}\label{f:SFE-rp}
\end{figure}

Here we compare the net SFE from our RP-only simulations with the
analytic predictions by \citet{ras16} and \citet{tho16}. The key
underlying assumptions of these models are that (1) the probability
density function (PDF) of gas surface density follows a log-normal
distribution, characteristic of supersonic isothermal turbulence and
that (2) only the gas with surface density below the Eddington surface
density is ejected. These analytic models predict a higher SFE for a
turbulent cloud than for an equivalent uniform cloud
\citep[e.g.,][]{fal10,kim16} with the same mass and size because a
greater luminosity is required to eject gas compressed to high surface
density by turbulence.

\citet{ras16} argued that for a cloud with given log-normal variance
in the surface density $\sigma_{\ln \Sigma^c}$, the luminosity would
continue to rise until it reaches a level that maximizes the outflow
efficiency. At this point, the final SFE would be bracketed between
two levels, $\varepsilon_{\rm *,min}$ and $\varepsilon_{\rm *,max}$.
At one extreme, all the remaining gas can be ejected without forming
stars, if the PDF adjusts itself rapidly to a successively lower peak
as gas is expelled. At the opposite extreme, all of the remaining gas
can turn into stars if it collapses before the PDF adjusts.
\citet{tho16} considered a similar situation, but they allowed for a
time-dependent star formation rate
$\dot M_* =\varepsilon_{\rm ff} M_{\rm gas}(t)/t_{\rm ff}$, where
$\varepsilon_{\rm ff}$ is a free parameter. In both models, gas is
assumed to be ejected rapidly.

In our simulations, the gas column density distribution remains broad
over the entire evolution. The column density PDFs closely resemble
log-normal functions with $\sigma_{\ln \Sigma^c} \sim 1.2$--$1.6$ that
do not vary much over the star-forming period (from $t_{*,10\%}$ to
$t_{*,90\%}$), roughly consistent with the results of \citet{ras16}.
The effective SFE per freefall time
$\varepsilon_{\rm ff,eff} = \SFE \tffo/(t_{*,99\%} - t_{*,0})$ in each
model turns out to vary in the range of $0.1$--$0.22$ across our
models. Both analytic models in principle allow for the cloud size to
vary over time, although the numerical simulations of \citet{ras16}
show that the effective radius in fact varies very little up until the
cloud is rapidly dispersed. Here, to compare to the analytic models,
we assume the cloud size is fixed at its initial radius ($x=1$ in
\citealt{ras16} and $p=0$ in \citealt{tho16}), and we take a fixed
value $\sigma_{\ln \Sigma^c} = 1.4$.

\autoref{f:SFE-rp} compares the net SFE resulting from our RP-only
runs (squares) with the theoretical predictions (various lines). For
this purpose, we adjust the numerical results, i.e.,
$\varepsilon_{\rm *,adj}= \SFE/(1-\varepsilon_{\rm ej,turb})$, to
allow for initial turbulent outflow, similar to \citet{ras16}. The
numerical results lie between $\varepsilon_{*,{\rm min}}$ and
$\varepsilon_{*,{\rm max}}$ (black solid lines) predicted by
\citet{ras16}. The net SFE is to some extent closer to
$\varepsilon_{*,{\rm min}}$ than $\varepsilon_{*,{\rm max}}$, which is
in contrast to the numerical results of \citet{ras16} (see their
Figure~25). In \citetalias{kim17}, we previously showed that the net
SFE obtained using the $M_1$-closure as in \citet{ras16} is higher
than that obtained from the adaptive ray tracing method, because in
the former the source function is smoothed out over a finite region
and thus the radiation force in the immediate vicinity of star
particles is lower than it should be, allowing additional accretion of
nearby gas.

\autoref{f:SFE-rp} plots as a dotted lines the net SFE predicted by
the \citeauthor{tho16} formalism, adopting $\varepsilon_{\rm ff}=0.1$
and $0.2$ to bracket our simulation results (note that they adopted a
much lower value for their fiducial model).  The prediction brackets
the results of our simulations.

We conclude that the results of our RP-only simulations are overall in
good agreement with the recent analytic models of \citet{ras16} and
\citet{tho16} for the net SFE in turbulent star-forming clouds
regulated by radiation pressure.


\bibliographystyle{apj}









\end{document}